\documentclass{aa}  

\usepackage[utf8]{inputenc}
\usepackage{graphicx}
\usepackage{dcolumn}
\usepackage{lineno}

\usepackage{xcolor}
\usepackage{bm}

\usepackage{color}
\usepackage{verbatim}
\usepackage{ulem}

\usepackage{amsmath}
\usepackage[hang]{footmisc}

\def\ztffid{$956$}
\def\sdssfid{$839$}
\def\psonefid{$785$}
\def\desfid{$905$}

\def\ztfcon{$386$}
\def\sdsscon{$606$}
\def\psonecon{$501$}
\def\descon{$556$}
\usepackage{txfonts}
\begin{document}

   \title{ZTF SN Ia DR2: Evidence of Changing Dust Distributions With Redshift Using Type Ia Supernovae }

    \author{B.~Popovic\inst{\ref{ip2i}}\fnmsep\thanks{Corresponding author: \texttt{b.popovic@ip2i.in2p3.fr}}
    \and M.~Rigault\inst{\ref{ip2i}}
    \and M.~Smith\inst{\ref{ip2i},\ref{lancaster}}
    \and M.~Ginolin\inst{\ref{ip2i}}
    \and A.~Goobar\inst{\ref{okc}}
    \and W.~D.~Kenworthy\inst{\ref{okc}}
    \and C.~Ganot\inst{1}
    \and F.~Ruppin\inst{\ref{ip2i}}
    \and G.~Dimitriadis\inst{\ref{dublin}},
    \and
    J.~Johansson\inst{\ref{okc}}
    \and M. Amenouche\inst{\ref{canada}}
    \and M.~Aubert\inst{\ref{clermont}} 
    \and C.~Barjou-Delayre\inst{\ref{clermont}} 
    \and U.~Burgaz\inst{\ref{dublin}}
    \and B.~Carreres\inst{\ref{marseille},\ref{duke}}
    \and F. Feinstein\inst{\ref{marseille}}
    \and D. Fouchez\inst{\ref{marseille}}
    \and L.~Galbany\inst{\ref{barcelona1},\ref{barcelona2}}
    \and T.~de~Jaeger\inst{\ref{lpnhe}}
    \and Y.-L.~Kim\inst{\ref{lancaster}},
    \and L.~Lacroix\inst{\ref{lpnhe}}
    \and P.~E. Nugent\inst{}
    \and B.~Racine\inst{\ref{marseille}},
    \and D.~Rosselli\inst{\ref{marseille}}
    \and P.~Rosnet\inst{\ref{clermont}}
    \and J.~Sollerman\inst{\ref{okc2}}
    \and D.~Hale\inst{11}
    \and R.~Laher\inst{\ref{caltecphysics}}
    \and T.~E.~M\"uller-Bravo\inst{\ref{barcelona1},\ref{barcelona2}}
    \and R.~Reed\inst{\ref{caltecphysics}}
    \and B.~Rusholme\inst{\ref{caltecphysics}}
    \and J.~Terwel\inst{\ref{dublin},\ref{not}}
    }
    \institute{
    Universite Claude Bernard Lyon 1, CNRS, IP2I Lyon / IN2P3, IMR 5822, F-69622 Villeurbanne, France
    \label{ip2i}
    \and
    Department of Physics, Lancaster University, Lancs LA1 4YB, UK \label{lancaster}
    \and
    Sorbonne Université, CNRS/IN2P3, LPNHE, F-75005, Paris, France
    \label{lpnhe}
    \and   
    The Oskar Klein Centre, Department of Physics, AlbaNova, SE-106 91 Stockholm , Sweden
    \label{okc}
    \and
    School of Physics, Trinity College Dublin, College Green, Dublin 2, Ireland
    \label{dublin}
    \and
    National Research Council of Canada, Herzberg Astronomy \& Astrophysics Research Centre, 5071 West Saanich Road, Victoria, BC V9E 2E7, Canada
    \label{canada}
    \and
    Université Clermont Auvergne, CNRS/IN2P3, LPCA, F-63000 Clermont-Ferrand, France
    \label{clermont} 
    \and
    DIRAC Institute, Department of Astronomy, University of Washington, 3910 15th Avenue NE, Seattle, WA 98195, USA
    \label{dirac}
    \and
    Aix Marseille Université, CNRS/IN2P3, CPPM, Marseille, France
    \label{marseille}
    \and    
    Department of Physics, Duke University Durham, NC 27708, USA
    \label{duke}
    \and   
    Institute of Astronomy and Kavli Institute for Cosmology, University of Cambridge, Madingley Road, Cambridge CB3 0HA, UK
    \label{cambridge}
    \and   
    Institute of Space Sciences (ICE-CSIC), Campus UAB, Carrer de Can Magrans, s/n, E-08193 Barcelona, Spain.
    \label{barcelona1}
    \and
    Institut d'Estudis Espacials de Catalunya (IEEC), 08860 Castelldefels (Barcelona), Spain
    \label{barcelona2}
    \and   
    Division of Physics, Mathematics \& Astronomy, California Institute of Technology, Pasadena, CA 91125, USA
    \label{caltecphysics}
    \and
    Deutsches Elektronen-Synchrotron DESY, Platanenallee 6, 15738 Zeuthen, Germany
    \label{desy}
    \and   
    Institut für Physik, Humboldt-Universität zu Berlin, Newtonstr. 15, 12489 Berlin, Germany
    \label{humbolt}
    \and   
    Caltech Optical Observatories, California Institute of Technology, Pasadena, CA 91125
    \label{caltecoptical}
    \and
    The Oskar Klein Centre, Department of Astronomy, Stockholm University, AlbaNova, SE-106 91 Stockholm , Sweden
    \label{okc2}
    \and
    Nordic Optical Telescope, Rambla José Ana Fernández Pérez 7, ES-38711 Breña Baja, Spain
    \label{not}
    }
    \date{\textit{in prep.}}

 
  \abstract
   {Type Ia supernova (SNIa) are excellent probes of local distance, and the increasing sample sizes of SNIa have driven an increased need to study the associated systematic uncertainties and improve the standardisation methods in preparation for the next generation of cosmological surveys into the dark energy equation-of-state $w$.}
   {We aim to probe the potential change in the SNIa standardisation parameter $c$ with redshift and the host-galaxy of the supernova. Improving the standardisation of SNIa brightnesses will require accounting for the relationship between the host and the SNIa, and potential shifts in the SNIa standardisation parameters with redshift will cause biases in the recovered cosmology. }
   {Here, we assemble a volume-limited sample of ~3000 likely SNIa across a redshift range of $z = 0.015$ to $z = 0.36$. This sample is fitted with changing mass and redshift bins to determine the relationship between intrinsic properties of SNe Ia and their redshift and host galaxy parameters. We then investigate the colour-luminosity parameter $\beta$ as a further test of the SNIa standardisation process.}
   {We find that the changing colour distribution of SNe Ia with redshift is driven by dust at a confidence of $>4\sigma$. Additionally, we show a strong correlation between the host galaxy mass and the colour-luminosity coefficient $\beta$ ($> 4\sigma$), even when accounting for the quantity of dust in a host galaxy. }
   {}


    \titlerunning{Evidence of Changing Dust Distributions with Redshift}
    \authorrunning{B. Popovic et al.}
   \maketitle
%

\section{Introduction}\label{sec:Introduction}

The discovery of the accelerating expansion of the universe \citep{Riess98, Perlmutter99} was made using standardised Type Ia supernovae (SNe Ia). The cause of this accelerating expansion is still an unsolved mystery, but the standard model of cosmology associates it to a cosmological constant $\Lambda$ or, more generically, to Dark Energy. The two decades between the discovery of the accelerating expansion and now have seen sample sizes of SNe Ia grow from tens of supernovae to thousands. Modern measurements of the Dark Energy equation-of-state parameter $w$  have statistical uncertainties on the order of $\sim0.02$ \citep{Brout22, DES5YR, Amalgame}, and this field of research will soon be dominated by the systematic uncertainties. The decreasing statistical uncertainties places a greater need on understanding the systematic uncertainties that inhere in measurements of cosmology made with SNe Ia. 

Here, we further improve on work that studies the correlation between the properties of SNe Ia and their host galaxies \citep{Sullivan10,Rigault13, Uddin17,Rigault18,Popovic21a}. 
Most cosmology analyses with SNe Ia use a variation of the SALT \citep{Guy10} framework to fit Type Ia supernova (SNIa) light curves and standardise their brightnesses. SALT uses two parameters to standardise the SNIa brightness: a stretch parameter $x_1$ that encompasses the luminosity dependence on light-curve duration and a colour parameter $c$ that describes the wavelength-dependent luminosity. Since \citealp{Kelly10,Sullivan10}, these two parameters have been joined by an ad-hoc correction, the `mass step', that accounts for the otherwise-unexplained observed correlation between the standardised brightness of the SNIa and the properties of their host galaxy.

Correlations between SNIa properties and their environment have always been studied in the context of SN cosmology, and early-on a correlation between the lightcurve stretch and the host parameters was demonstrated \citep[e.g.][]{Hamuy96, Howell07, Sullivan10, Lampeitl10}. Recently, with increased statistical power, progress has been made into uncovering underlying reasons for the correlations between fitted SALT properties (e.g., $x_1$, $c$) and the properties of the host galaxies of the SNe Ia. While \cite{Popovic21a} found strong evidence that both $x_1$ and $c$ change with the host galaxy mass, \cite{Nicolas21} provided an analytical description of changing $x_1$ values with both host galaxy mass and Local specific Star Formation Rate (LsSFR). However, this refined description was unable to explain the mass step.
In the last decades, several explanations arose to find the origin of the mass-step, with the aim to correctly account for this SN astrophysical dependency while fitting cosmological parameters. \cite{Rigault13,Rigault18} and \cite{Briday21}, following earlier work on SN rates \citep[e.g.][]{Sullivan06} suggest that prompt and delayed SNe~Ia have intrinsically different absolute magnitudes. In that model, the prompt SNe, associated with recent star formation, are fainter, and the mass-step originates from the fact that massive galaxies host few prompt SNe~Ia (see also e.g. \citealt{Childress13,Childress14}).

Recently, \cite{BS20}, following work from e.g., \cite{Mandel17}, provided an explanation disconnected from the prompt and delayed rate-model. They show that the amplitude of the mass-step depends on the SN colour and were able explain the variation, and consequently the mass-step, by allowing the colour-magnitude relation to change as a function of host-mass. This model, further developed in \cite{Popovic22}, \cite{Kelsey22} and \cite{Wiseman22}, follows the assumption that the redder end of the SN~Ia color distribution is caused by dust, which properties varies as a function of host stellar mass. This dust model well describes the data \citep{Popovic22}.

While both approaches (dust or age) have their benefits and their limitations, to unveil the true origin of the observed astrophysical biases is of paramount importance for SN cosmology as the resulting modeling for deriving distances may impact the measurement of cosmological parameters.
In this analysis, we will use the second data release of the Zwicky Transient Facility (ZTF, \citealp{bellm19,graham19, masci19}) SNIa sample (\textcolor{red}{ZTF Data Release Paper}), with spectra from \cite{SEDm, IFU}. This release (ZTF-Cosmo-DR2) has $\sim3000$ cosmological quality SNe~Ia at $z<0.1$. Combined with recent improvements in the reliability of non-spectroscopically confirmed SNIa samples \citep{DES5YR, Amalgame}, we present the first rigorous study of the evolution of SNIa colour and colour-magnitude relation with redshift.

{This study is important for the future of studies of the equation-of-state $w$. Combining constraints from SNe Ia with external probes provides exquisite measurements of the properties of our universe, as in \cite{DES5YR} and \cite{DESIY1}. Systematics that evolve with redshift are degenerate with different cosmologies. With recent results from \cite{DESIY1} that indicate a non-static $w$ may be preferred, even-further increased control of systematic uncertainties will be needed for the next generation of measurements of cosmology with SNe Ia.}

The overview of the paper is as follows: Section \ref{sec:Data} provides an overview of the data: ZTF, SDSS, PS1, and DES, alongside quality cuts and light-curve fitting parameters. Section \ref{sec:Methodology} explains the methodology used to define and fit the sample to explore evolution of $c$ with redshift and mass. Section \ref{sec:Results} presents the results and any potential evolution of $c$ with other SNIa parameters, while Sections \ref{sec:Discussion} and \ref{sec:Conclusion} host the discussion and conclusions respectively.

\section{Data}\label{sec:Data}

{High and low redshift surveys require different cadences, sky-coverage, and overall survey strategies. In order to maximise statistics and redshift range, samples of SNIa are often combined together, such as in \cite{Union, Betoule14, Brout22, Amalgame}.}

We make use of the SNIa samples from the Zwicky Transient Facility (ZTF; \textcolor{red}{ZTF Data Release Paper}), the Sloan Digital Sky Survey (SDSS; \citealp{Frieman08, Sako18, Popovic19}), Pan-STARRS (PS1; \citealp{Jones18}), and Dark Energy Survey 5-year (DES; \citealp{DES5YR}). Table~\ref{tab:NUMBERS} provides a summary of the dataset used in this analysis, showing the magnitude limit of the telescope, the resulting redshift cut, and the number of SNe in the final sample. We include an additional conservative sample, with a stricter redshift cut, that is shown in parentheses. This is explained further in Section \ref{sec:Methodology}.

\begin{table}[]
    \centering
    \begin{tabular}{c|c|c|c}
        Survey & mag$_{\rm lim}$ & Fid $z_{\rm lim}$ (Con) & Fid $N_{\rm SN}$ (Con)  \\
        \hline
         ZTF  & 18.75 & 0.06 (0.04) & \ztffid (\ztfcon) \\
         SDSS & 22.5 & 0.24 (0.20) & \sdssfid (\sdsscon) \\
         PS1  & 23.1 & 0.30 (0.25) & \psonefid (\psonecon) \\
         DES  & 23.5 & 0.36 (0.30) & \desfid (\descon) \\
           &  & &  \\
         Total & -- & -- & {3485 (2049)}
    \end{tabular}
    \caption{Summary of redshift cuts for Fiducial (Fid) and Conservative (Con) samples {after other quality cuts.}}
    \label{tab:NUMBERS}
\end{table}

The second ZTF data release covers the years of 2018 to 2020. The survey overview is presented in \textcolor{red}{ZTF Data Release Paper}. Lightcurve data typical have a 2 day cadence in $g$ and $r$ and a 5 day cadence in $i$. All SNe~Ia have a spectroscopic classification and about half have a host-galaxy redshift with the typical precision of $10^{-4}$. For the other half, the data release relies on SN spectroscopic feature that has been shown to have an unbiased redshift precision of $10^{-3}$. {Table \ref{tab:DataSum} shows a summary of the ZTF sample compared to the largest historic low-redshift surveys of SNe Ia; while all cover a similar redshift range ($z<0.1$), ZTF is the first all-sky survey with statistics an order of magnitude larger than previous samples.}

\begin{table}[]
    \centering
    \begin{tabular}{c|ccc}
        Survey & Strategy & $N_{SN}$ & Paper \\
        \hline
        ZTF & All-sky & 2628 & \textcolor{red}{ZTF Data Release Paper} \\
        Foundation & Untargeted & 176 & \cite{Foley18} \\
        CSP & Targeted & 138 & \cite{Krisciunas17}\\
        CfA & Targeted & 185 & \cite{Hicken09a}
    \end{tabular}
    \caption{{Comparison of ZTF with historical low-redshift samples}}
    \label{tab:DataSum}
\end{table}

The SDSS supernova program ran for 3 observing seasons between 2005 to 2007. The survey overview is provided in \cite{Frieman08}. SDSS lightcurves have  data in the \textit{ugriz} bands with an average observing cadence of 4 days. We take light curves from \cite{Sako18} and host galaxies and redshifts from \cite{Popovic19}.

PS1 ran from 2009 to 2013, observing with the \textit{griz} filters at a cadence of 6 observations every 5 days. The PS1 light curves are taken from \cite{Chambers16}, with host galaxy redshifts drawn from papers listed in \cite{Jones18}.

The DES-SN program ran for 5 years, covering 23 deg$^2$ across the sky. With an average observing cadence of 7 days, DES took data in the \textit{griz} bands across eight "shallow" fields and two "deep" fields, covering a balance between increased maximum redshift in the deep fields, which utilise multiple exposures, and area coverage and volume in the shallow fields; the data comes from Sanchez et al. 2024.

SNe Ia at peak brightness are not uniform enough to support accurate measurements of cosmological parameters on their own; they require standardisation via a light-curve modeling and fitting program. We use the SALT2 program introduced by \cite{Guy10} with updated training from \cite{Taylor21}. The SALT fit returns four parameters for each SNIa: $t_0$ the time of peak brightness; $x_0$ the overall light-curve amplitude; $c$ the colour parameter; and $x_1$, the stretch parameter related to the width of the light-curve. These latter three parameters are directly used to standardise SNIa luminosity $\mu$ according to the Tripp estimator \citep{Tripp98}:
\begin{equation}\label{eq:tripp}
\mu = m_B + \alpha x_1 - \beta c - M_0  
\end{equation}
where $m_B = -2.5\log_{10}(x_0)$, $c$ and $x_1$ are defined above, and $M_0$ is the absolute magnitude of a SNIa with $c = x_1 = 0$. $\alpha$ and $\beta$ are the stretch-luminosity and colour-luminosity coefficiencts respectively, and are defined for a given sample of SNIa.

For the data, we apply the following conventional cuts, following papers such as \cite{Betoule14, Scolnic18, Brout22}:
\begin{itemize}
    \item $\sigma_{x_1} < 1$ : SALT2 $x_1$ uncertainty $< 1$.
    \item  $\sigma_{c} < 0.1$ : SALT2 $c$ uncertainty $< 0.1$
    \item $\sigma_{t_0} < 1$ : Uncertainty on fitted peak brightness epoch $< 1$ days.
    \item $-4 < x_1 < 4$.
    \item $-0.3 < c < 0.8$.
    \item $T_{\rm rest, min} < 5$ : Requires at least 1 observation 5 days before peak brightness (rest frame).
    \item $T_{\rm rest, max} > 0$ : Requires at least 1 observation after peak brightness (rest frame).
\end{itemize}

\begin{figure}[!h]
    \centering
    \includegraphics[width=8cm]{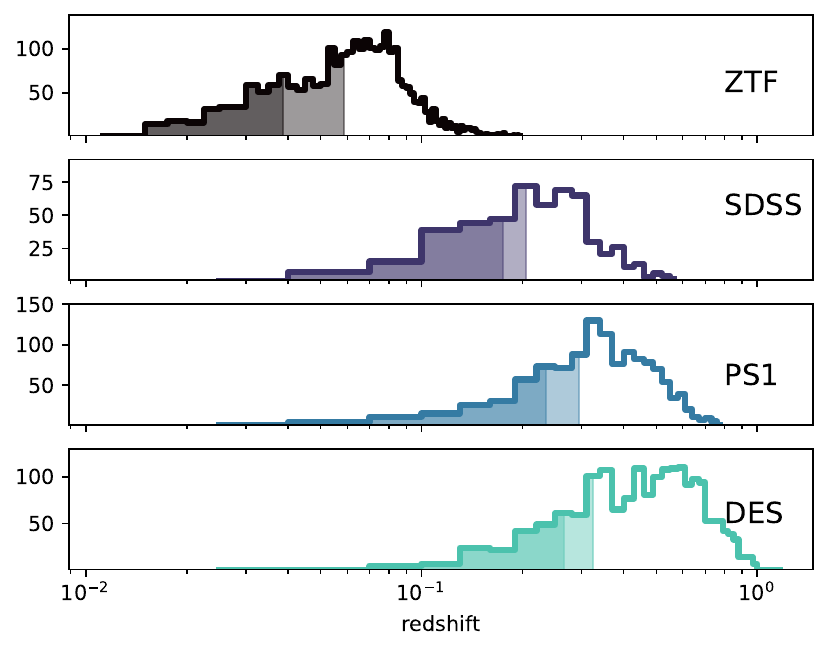}
    \caption{The redshift distributions for ZTF, SDSS, PS1, and DES. The full sample is presented for each survey before any redshift cuts in unfilled histogram. The light fill represents the redshift range of the fiducial sample, and the dark fill shows the redshift range of the conservative sample.}
    \label{fig:Redshifts}
\end{figure}

In the case for SDSS, PS1, and DES, we institute an additional cut, that the probability of a SN being a SNIa is greater than 0.9: $P_{Ia} > 0.9$. We make use of the SuperNNova classifier from \cite{Moller19}, taking trained surfaces from \cite{Amalgame} and \cite{DES5YR}. This cut is not applied for ZTF, which has spectroscopic identification for the entire sample. {The extra cut placed on the high redshift surveys ensures that the sample is comprised of SNe Ia, and that the limiting magnitude of these surveys is determined by the survey photometry rather than the spectroscopic followup, bringing them in line with the ZTF sample.}

\section{Methodology}\label{sec:Methodology}

{This paper is made possible by the unprecedented redshift range and statistics of the ZTF sample. Combining ZTF with the higher-redshift surveys -- SDSS, PS1, DES -- results in a volume-limited sample of $\sim3500$ SNe Ia ranging from $0.04 < z < 0.36$, larger than any previously assembled collection of SNe Ia.}

Here we outline the methodology used to assess the evolution of SNIa colour with redshift and host galaxy mass. In this section, we will begin with the definition and creation of a volume-limited sample for each survey so as to remove potential biases arising from selection effects, then split the samples into redshift and mass bins to fit the colour population within each bin.

A consequence of the need for standardisation for SNIa is that redder SNIa (those approaching $c = 0.3$) are dimmer than bluer supernovae. This presents a problem for investigating the colour evolution with redshift; as redshift increases, surveys approach the magnitude limit of being able to observe objects. This Malmquist bias can impact inferred cosmologies, and indeed our colour distributions. 

\begin{figure*}[!t]
    \centering
    \includegraphics[width=16cm]{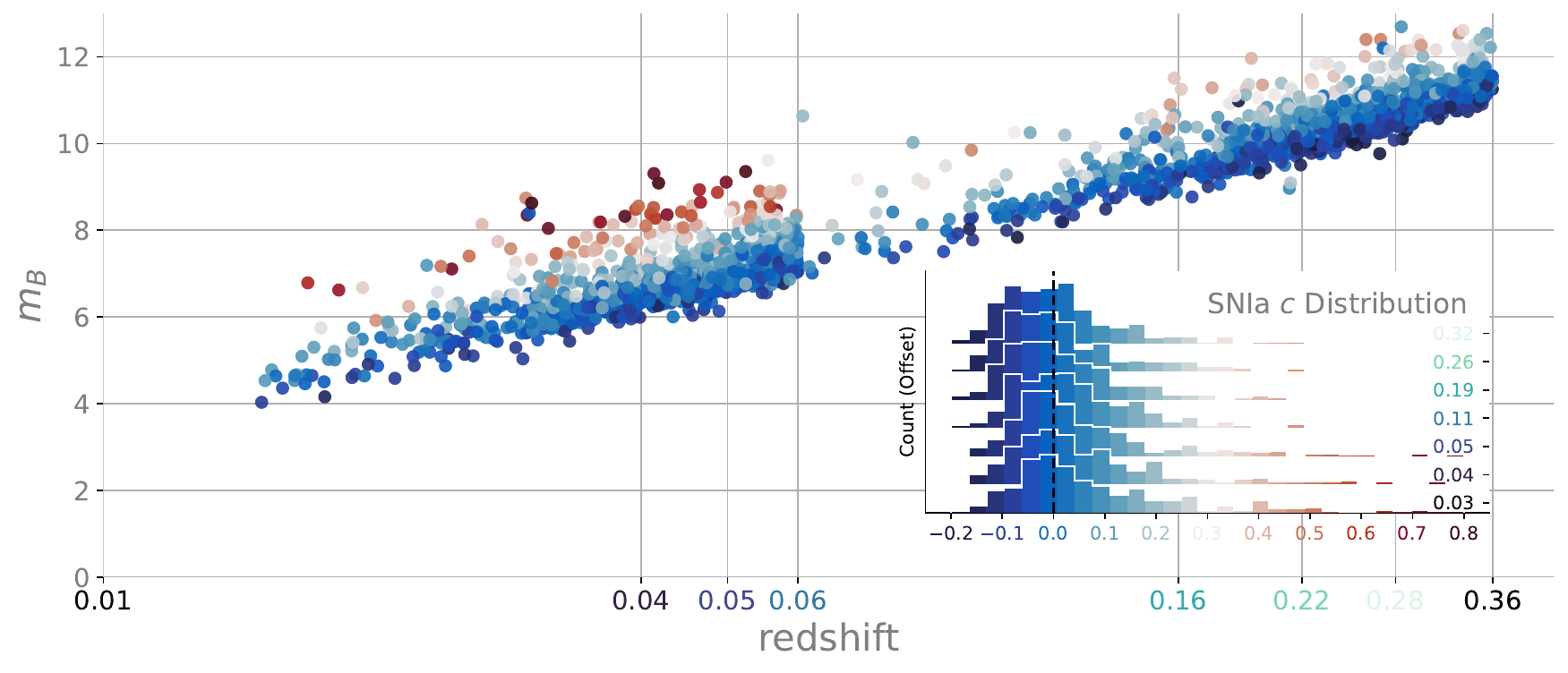}
    \caption{$m_B$ vs redshift for the SNIa in the Fiducial Sample defined in Table \ref{tab:NUMBERS}. Each supernova is colour coded according to its colour $c$. The grey  vertical lines demark the quantile redshift bins, and the colour histogram is provided for each redshift bin in the inset plot. }
    \label{fig:MAINPLOT}
\end{figure*}

There are two solutions to this magnitude limit. The first is the introduction of bias corrections from \cite{Kessler16}, using realistic simulations to correct for limitations in observing the apparent magnitudes of SNIa. The second method is to create a `volume limited' sample. Since the purpose of this work is to find the underlying distribution of colour, we are not able to provide accurate bias corrections. Instead, we {follow \cite{Nicolas21}} and define two volume limited samples for each survey: a fiducial survey, and the conservative one.

To define the volume limited sample, we follow \cite{Nicolas21}. {We follow Equation \ref{eq:tripp} and assume that the absolute magnitude for an SNIa with $c=x_1=0$ is $M_0 -19.36$. From this, we can calculate a brightness at time of maximum light $M = M_0 - \alpha x_1 + \beta c$. Assuming a common $\beta = 3.1$ and $\alpha = 0.15$ \citep{Betoule14, Scolnic18, Brout22}, the faintest magnitude we would expect to see is approximately $-18.31$ mags, from a SNIa with $x_1 = -1.65$ and $c = 0.25$. However, we require at least one observation \textit{before} peak brightness, which requires an even fainter magnitude, which we place at $-18$ for our fiducial limiting magnitude. Our conservative limiting magnitude we place fainter, at $-17.5$. In both cases, the limiting magnitudes we place are in reference to the SALT-fitted brightness.}

With the limiting magnitudes, and the magnitude limit of the surveys (taken from \citealp{Nicolas21}, \textcolor{red}{ZTF Data Release Paper}), we are able to calculate a redshift $z_{\rm lim}$ such that we expect the sample to be complete and free of magnitude-based selection effects below $z_{\rm lim}$. A review of the magnitude limits, limiting redshifts, and resulting number of SNIa is shown for each survey in Table \ref{tab:NUMBERS}. The redshift distributions: full, fiducial, and conservative; is shown in Figure \ref{fig:Redshifts} for each survey in this paper. {We place our redshift cuts on the sample after calculating the limiting redshift $z_{\rm lim}$ from the SALT-fitted magnitude of the supernovae.}

To investigate evolution of SNIa colour $c$ with redshift, we {quantile-bin our sample to create 7 evenly-populated bins of redshift.} For colour evolution with host galaxy stellar masses measured relative to the mass of our sun (Sol = 1 $M_{\odot}$, $M_{\rm Star} = 10^X M_{\odot}$, where here we report $X$), we make use of evenly-spaced mass bins, so as to maintain a bin edge at $M_{\rm Star} = 10$, to be consistent with literature involving the mass step at $10 M_{\rm Star}$.

In both cases, we fit within the individual bins of mass or redshift separately to find the apparent colour distribution. The change of the fit results in each bin can be assessed to determine any evolution.

Within the individual bins, we make use of the fitting code from \cite{Ginolin24b} to fit the apparent colour via a convolution of a Gaussian intrinsic SNIa colour $c_{\rm int}$ and an exponential dust component $E_{\rm dust}$:
\begin{equation}
    c_{\rm obs} = c_{\rm int} + E_{\rm dust} + \epsilon_{\rm noise}
\label{eq:cobs}
\end{equation}
with a noise term $\epsilon_{\rm noise}$. The Gaussian and exponential tail is drawn as such:
\begin{equation}
    P(c)= \mathcal{N}(c\,|\,\overline{c}, c_{\sigma}) \ast \begin{cases} 0 & \mbox{if } c \leq 0 \\ \frac{1}{E_{\rm dust}}e^{-c/E_{\rm dust}} & \mbox{if } c > 0
    \end{cases}
    \label{eq:cplusdust}
\end{equation}
The minimisation process is done via \texttt{MINUIT} \citep{MINUIT}, which also returns fitting errors that are used within the analysis. {The errors recorded by \texttt{MINUIT} are symmetric Gaussians}. We report these results as $\overline{c}$, the mean of the Gaussian intrinsic colour distribution, $c_{\sigma}$, the standard deviation, and $E_{\rm Dust}$, the $\tau$ value describing the exponential distribution. 

{The choice here of an intrinsic Gaussian distribution that is reddened by external dust is taken from \cite{BS20}, but has recently gained traction with SNIa cosmology, used in works such as \cite{Wiseman22, Kelsey22, Popovic21a} and as the default model in the cosmology analyses performed by \cite{Brout22, DES5YR, Amalgame}.}

\cite{BS20} and \cite{Popovic22} showed that while the observed SNIa colour distribution is an acceptable probe of the columnar density of dust $E_{\rm Dust}$ (e.g. the quantity of dust through the line-of-sight), it is not able to provide strong constraints on the $R_v$ parameter. The best indicator of the $R_v$ of the host galaxy, from SNIa data alone, is the colour-luminosity coefficient $\beta$. To further investigate the relationship between colour and host-galaxy properties, we investigate the relationship between host galaxy mass and the $\beta$ parameter.

To fit $\beta$, we modify Equation \ref{eq:tripp}:
\begin{equation}\label{eq:trippmod}
    \mu^* = m_B - \mu_{\rm theory} + \alpha x_1
\end{equation}
where $m_B$ and $x_1$ are as normally defined, and we float $\alpha$ alongside our $\beta$. $\mu_{\rm theory}$ is the theoretical distance modulus from a Flat $\Lambda$CDM cosmology with $H_0 = 74$. {The slope of this $\mu^*$ value, when plotted against the supernova colour $c$, is $\beta$.} We fit this slope, simultaneously with the intercept, using \texttt{MINUIT} in bins of mass to determine potential $\beta$ evolution.

The masses for this paper are taken from their respective data releases, e.g. \textcolor{red}{ZTF DR2 Data Release} and \cite{DES5YR} for ZTF and DES, or in the case of SDSS and PS1, the updated masses from \cite{Amalgame}. {All surveys} use consistent PEGASE.2 code with a \cite{Kroupa01} initial mass function. 

\section{Results}\label{sec:Results}

We present the results of the fitting process from Section \ref{sec:Methodology} on the samples defined in Section \ref{sec:Data}. Figure \ref{fig:MAINPLOT} shows $m_B$ vs. redshift for the fiducial sample, alongside the demarcations for the quantile redshift bins. {In inset, we show the colour distributions of each of the redshift bins; here we will investigate the cause of the apparent change in colour distribution.} Significance measurements are calculated by comparing the binned results to the mean of the overall distribution, and converting to a significance measurement using the number of degrees of freedom. 

\subsection{Evolution With Redshift}\label{sec:Results:subsec:Redshift}

\begin{figure}
    \centering
    \includegraphics[width=8cm]{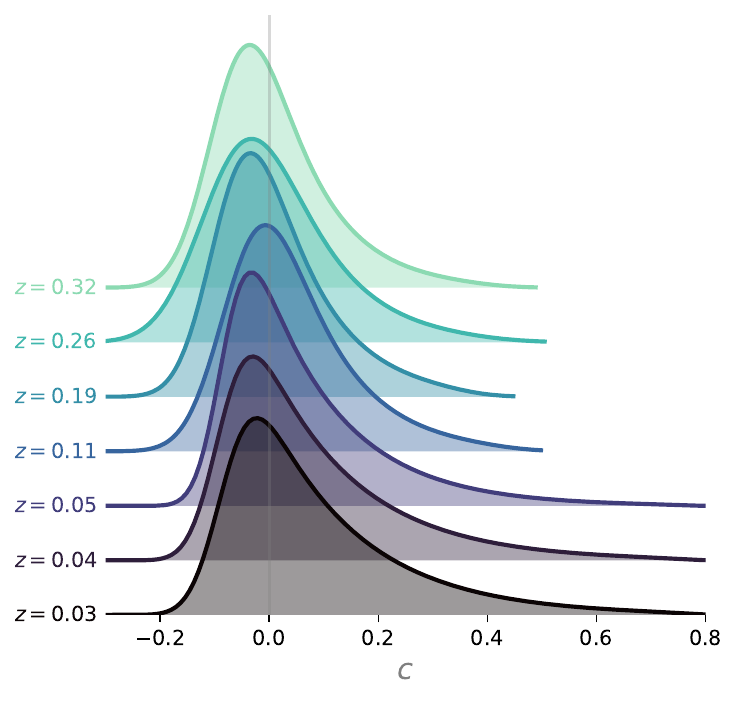}
    \caption{Fit results for the quantile redshift bins of the fiducial sample. The $c$ distribution is shown for each redshift bin, colour coded to become darker with increasing redshift.}
    \label{fig:CFITQUANT}
\end{figure}

We fit the colour distribution of the SNIa in each of the 7 quantile redshift bins and present the inferred Gaussian+exponential distribution in Figure \ref{fig:CFITQUANT}. 

\begin{figure}[h]
    \centering
    \includegraphics[width=8cm]{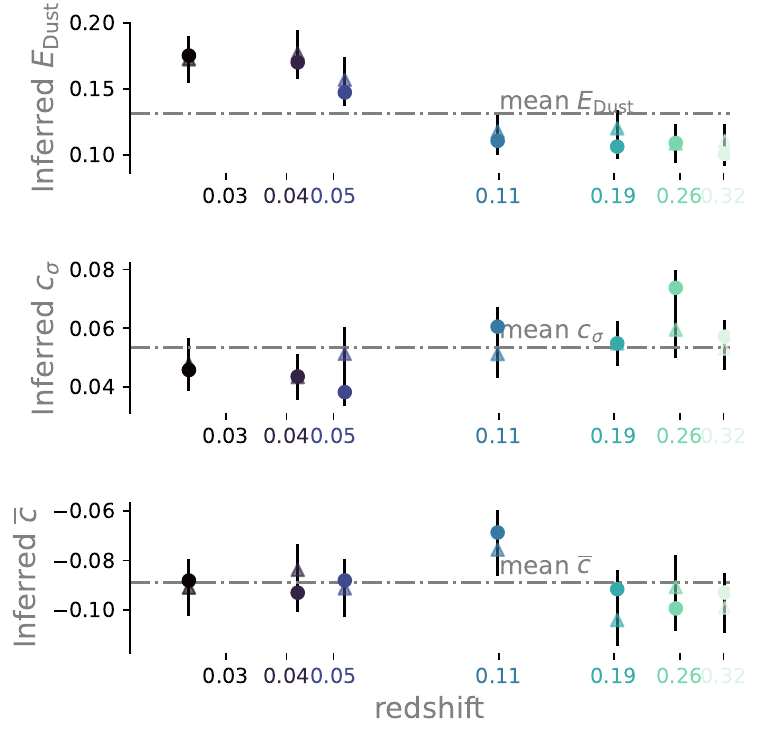}
    \caption{The fit parameters for the Gaussian and exponential distributions plotted with increasing redshift. Alongside each parameter, the mean fit value is plotted in dash-dotted line to provide context. The fit results from the conservative sample are overlaid in triangles for comparison.}
    \label{fig:PLOTEVOQUANT}
\end{figure}

We plot the Gaussian and exponential parameters from Figure \ref{fig:CFITQUANT} in Figure \ref{fig:PLOTEVOQUANT} to more accurately assess potential trends. We find that the Gaussian parameters that describe the SNIa colour distribution are consistent with no evolution. $\overline{c}$ changes with redshift at a confidence of $<1\sigma$, and $c_{\sigma}$  changes with redshift at a confidence of $3.4\sigma$. These low signals are in contrast with the $E_{\rm Dust}$ parameter, for which we find evidence of evolution with redshift at a confidence of $>6\sigma$. For comparison, we show the $\chi^2$ values from the conservative sample that was detailed in Section \ref{sec:Methodology}. The conservative sample trends are qualitatively similar to the full sample, though the larger uncertainties that arise from the smaller sample decrease the evidence of $E_{\rm Dust}$ evolving from $6\sigma$ to $3\sigma$.

Table \ref{tab:zchi2s} shows the $\chi^2$ breakdown for each of the parameters and their evolution with redshift. We see the trends from the fiducial sample -- data that prefers an evolving dust-based $E_{\rm Dust}$ parameter over a change in the intrinsic SNIa colour distribution -- holds for the conservative sample, though we find a confidence of $>5\sigma$ for $E_{\rm Dust}$ evolution in the conservative sample rather than $>6\sigma$.

\begin{table}[]
    \centering
    \begin{tabular}{c|c|c}
        Parameter & Fiducial $\chi^2$ & Conservative $\chi^2$  \\
         \hline
         $\overline{c}$ & 6.9 ($<1 \sigma$) & 4.7  ($<1\sigma$) \\
         $c_{\sigma}$ & 26.9 ($3.4\sigma$) & 2.1 ($<1\sigma$) \\
         $E_{\rm Dust}$ & 50.27 ($>6\sigma$) & 21.8  ($3\sigma$)
    \end{tabular}
    \caption{Goodness-of-fit $\chi^2$ values for each parameter compared to the null hypothesis of no $z$-evolution, with 7 DOF.}
    \label{tab:zchi2s}
\end{table}

\subsection{Evolution With Mass}\label{sec:Results:subsec:Mass}

\begin{figure}[h]
    \centering
    \includegraphics[width=8cm]{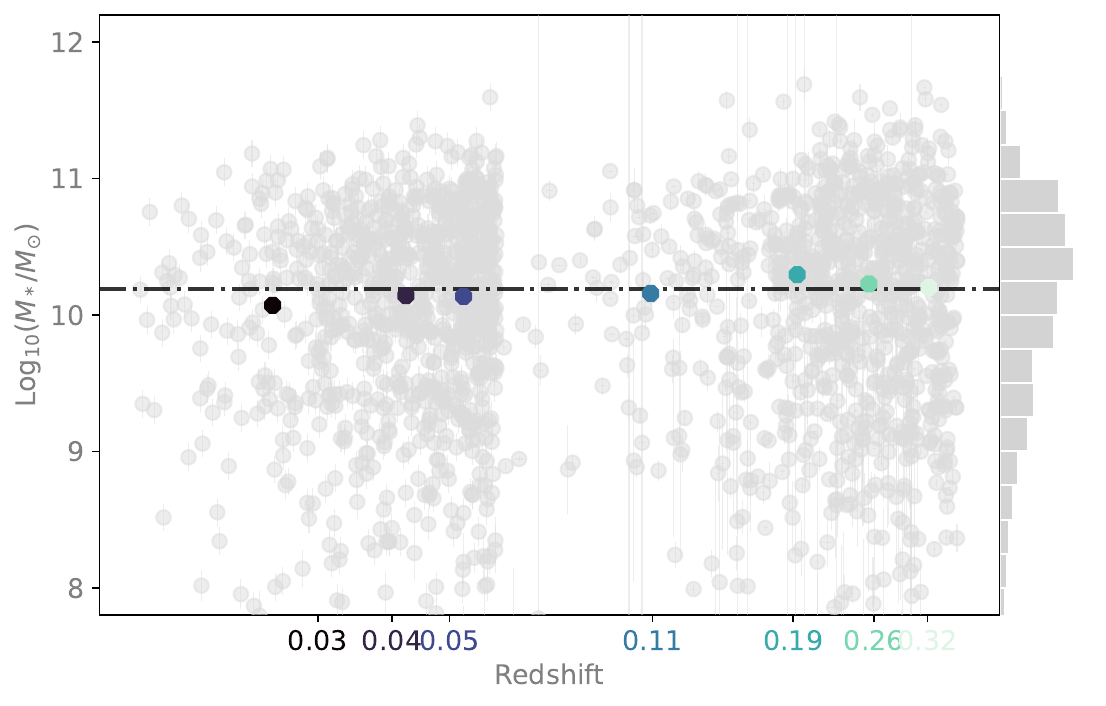}
    \caption{Host galaxy mass plotted vs redshift for the Fiducial sample. Supernovae are shown in grey, the coloured circles represent the median host galaxy mass in each quantile redshift bin. The dash-dotted black line shows the median host galaxy mass of the entire sample. The mass distribution, marginalised over redshift, is shown on the right.}
    \label{fig:MASSVREDSHIFT}
\end{figure}

Figure \ref{fig:MASSVREDSHIFT} shows the evolution of the host galaxy mass with redshift. There is an under-density of SNe between $z = 0.06$ and $z = 0.1$; {this under-density arises from the redshift cut on ZTF at $z=0.06$, leaving only SDSS SNe in the range of $0.06 < z 0.1$}. Even with this under-density we find no evidence of the median host galaxy mass changing with redshift {($<1\sigma$ from the median of the full distribution).} 

\begin{figure}[h]
    \centering
    \includegraphics[width=8cm]{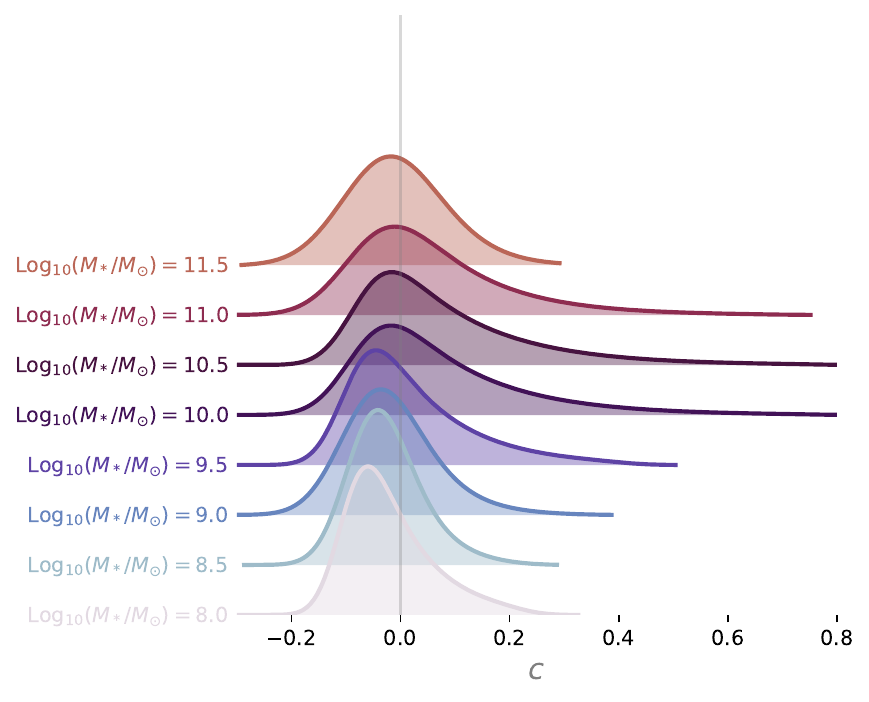}
    \caption{Fit results for the host galaxy mass bins of the fiducial sample. The $c$ distribution is shown for each mass bin, colour coded with increasing host galaxy mass.}
    \label{fig:CMASS}
\end{figure}

With confidence that there is no correlation between the host galaxy mass and redshift, we show the evolution of the SNIa colour distribution with host galaxy mass. Figure \ref{fig:CMASS} shows the change of colour distribution with host galaxy mass, binned in steps of $0.5 M_{\rm Star}$, starting at $8 M_{\rm Star}$ and going to $11.5 M_{\rm Star}$. As a reminder, in contrast to the quantile binning for redshift, we do not require equal statistics in each bin, so that we can place a bin edge at the historical mass step value of $10 M_{\rm Star}$. 

\begin{figure}[h]
    \centering
    \includegraphics[width=8cm]{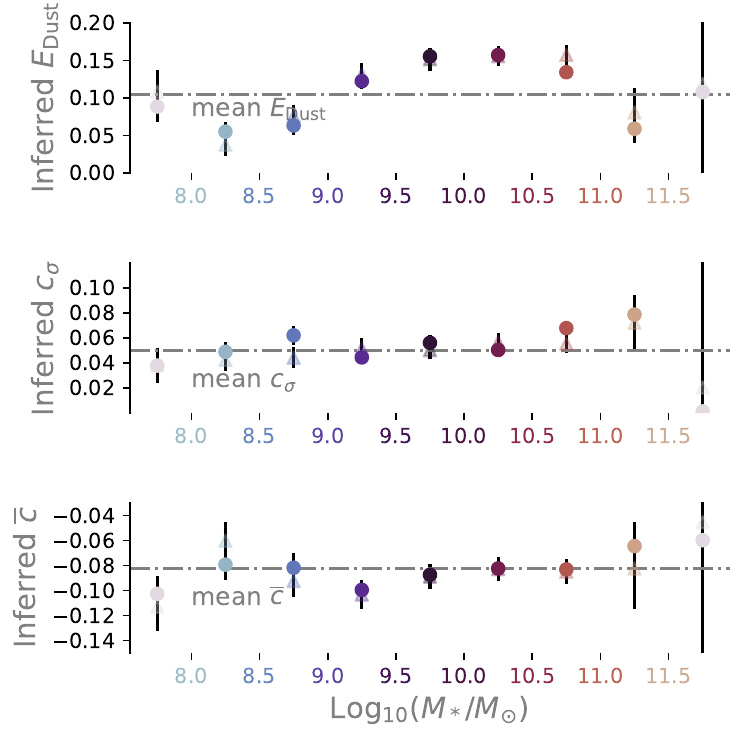}
    \caption{The fit parameters for the Gaussian and exponential distributions plotted with increasing host galaxy mass. Alongside each parameter, the mean fit values is plotted in dash-dotted line to provide context. The fit results from the conservative sample are overlaid with triangles for comparison.}
    \label{fig:MASSEVORESULTS}
\end{figure}

Figure \ref{fig:MASSEVORESULTS} displays the individual parameters $\overline{c}$, $c_{\sigma}$, and $E_{\rm Dust}$ from the fitted results presented in Figure \ref{fig:CMASS}. We show the $\chi^2$ values in Table \ref{tab:mchi2s}, and see results similar to the redshift distribution results. There is little evidence of the SNIa colour parameters changing, as $\overline{c}$ exhibits $<1\sigma$ change and $c_{\sigma}$ exhibits a $3.3\sigma$ signal of evolution with host galaxy mass. Again, the conservative sample, which is plotted alongside the fiducial, does not show significantly different trends, shown in Table \ref{tab:mchi2s}.

On the other hand, the $E_{\rm Dust}$ parameter shows a clear evolution, $>6\sigma$, with the mass of the host galaxy. The $E_{\rm Dust}$ values peak around the host galaxy mass of 10 $/M_{\rm Star}$, decreasing beyond this point for both higher and lower mass galaxies. The first mass bin, however, does not follow this trend. This does not appear to be a binning effect, as it persists across different bin edges and sizes. However, the discrepancy here may be caused by the mis-association of host galaxies, or biases arising from too-faint hosts \citep{Alfonso22}. This trend is equally strong with the conservative sample, at $>6\sigma$.

\begin{figure*}
    \centering
    \includegraphics[width=18cm]{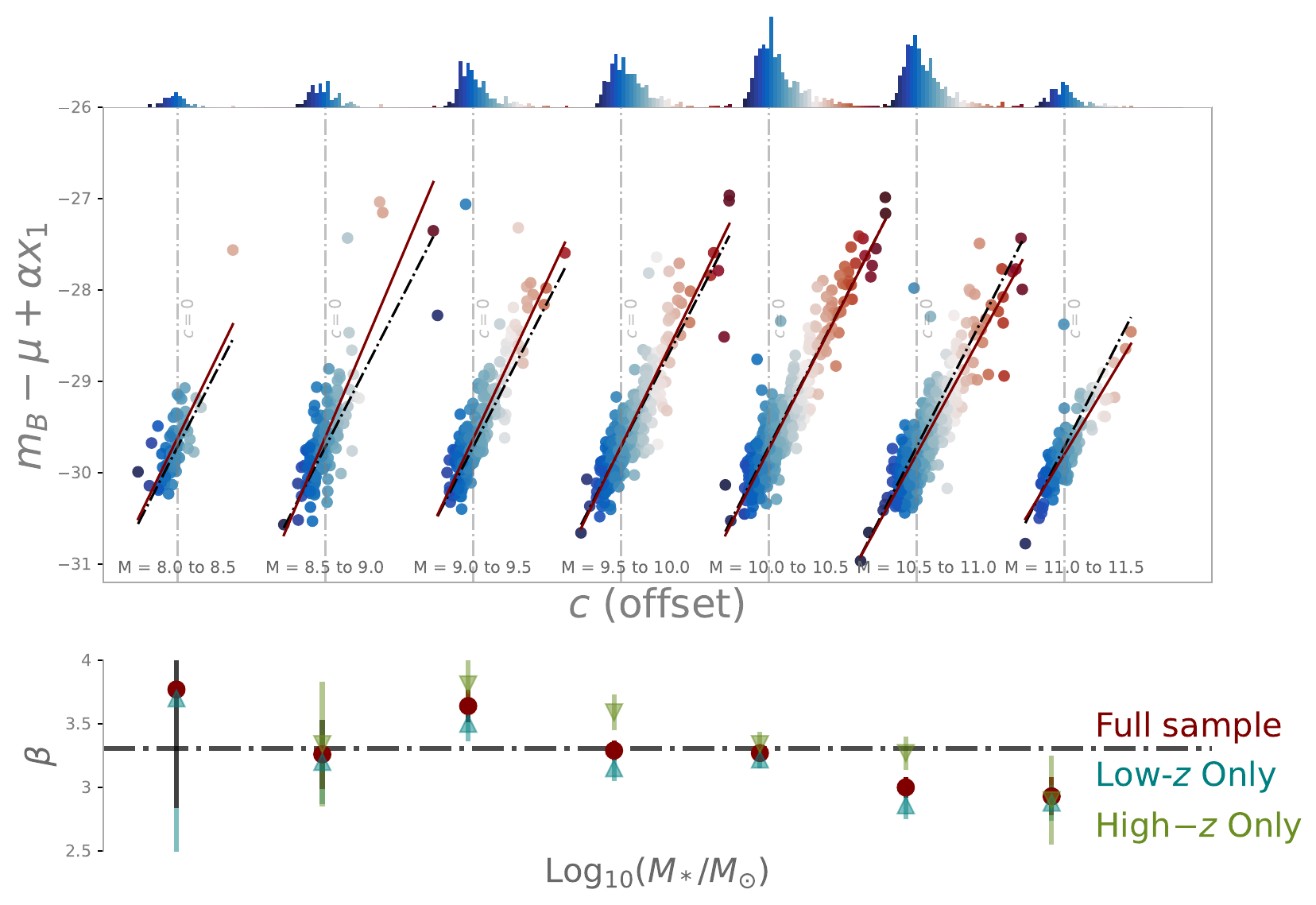}
    \caption{Main Plot: $c$ vs. $m_B - \mu + \alpha x_1$, offset in bins of host galaxy mass. The SNe are colour-coded by their colour $c$ for visual guidance. The slope ($\beta$ value) for each mass bin is shown in red, alongside the fiducial beta in black dash-dotted line. The light grey dash-dotted vertical line shows the $c=0$ point for each mass bin. Top Plot: Colour histograms for each of the mass bins is shown, colour-coded according to $c$. Inset: The $\beta$ vs mass relationship for the fitted mass bins. The $\beta$ values for the full sample are presented in brown points, the ZTF-only $\beta$ values in teal triangle, and the high-redshift sample in olive triangle, with the fiducial $\beta$ shown in grey dash-dotted line. }
    \label{fig:BetaVMass}
\end{figure*}

\begin{table}[]
    \centering
    \begin{tabular}{c|c|c}
        Parameter & Fiducial $\chi^2$ & Conservative $\chi^2$  \\
         \hline
         $\overline{c}$ & 9.4 ($<1\sigma$) & 8.4 ($<1\sigma$)  \\
         $c_{\sigma}$ & 27.4 ($3.3\sigma$) & 4.7 ($<1\sigma$)  \\
         $E_{\rm Dust}$ & 102 ($>6\sigma$) & 62.7 ($>6\sigma$)
    \end{tabular}
    \caption{Goodness-of-fit $\chi^2$ values for each parameter compared to the null hypothesis of no stellar mass evolution, with 9 DOF.}
    \label{tab:mchi2s}
\end{table}

For both the $z$ and host galaxy mass evolution tests, we attempt to account for potential non-Ia contamination in the signal by making use of simulations from \cite{Amalgame}. While these simulations are only available for SDSS and PS1, they are a good test of potential signal contaminant by representing the two surveys with the highest potential non-Ia contamination. We perform the redshift and host galaxy mass analysis two times, one with a simulated contamination and one with a pure-Ia sample. {We apply the same $P_{Ia} > 0.9$ cut to the contaminated simulated sample as we do to the real data, to test any bias due to this additional cut. This is compared with the pure SNIa simulated sample. }
We find no significant difference $(<0.5\sigma)$ in recovered results between the contaminated and un-contaminated samples. 

\subsection{Evolution of $\beta$ with host galaxy mass}

While the observed $c$ distribution is a tracer of $E_{\rm Dust}$, it does not provide strong constraints on the type or size of the dust particulates, conventionally denoted as $R_v$. Instead, $R_v$ has been shown to be better tracked with supernova data by the colour-luminosity relationship $\beta$ \citep{BS20, Popovic22, Johansson20}. Here we investigate the evolution of $\beta$ with the host galaxy mass, to see if there exists a change in dust properties alongside the observed changes in columnar density. 

Figure \ref{fig:BetaVMass} shows the colour-luminosity relationship for 7 mass bins across the fiducial sample. Again, we have fit $\alpha$ alongside our $\beta$ and assumed a Flat $\Lambda$CDM cosmology in making these plots. We focus on the inset that shows the fitted $\beta$ values for each mass bin. With the exception of the first bin, there exists a clear trend of decreasing $\beta$ with increasing host galaxy mass. Considering the bins that are populated for all three redshift categories (`Full', `Low-$z$', and `High-$z$'), we find evidence of a change in $\beta$ with mass at the $3\sigma$ level, likely driven by the lowest mass bins. These results are in line with those of \cite{Popovic22, Chen22} and provide further hints of changing host-galaxy properties. We discuss the inclusion of the first bin in Section \ref{sec:Discussion}.

\section{Discussion}\label{sec:Discussion}

\subsection{Redshift and Mass Evolution}

We find no strong evidence {for an evolving dust-free SNIa colour distribution with redshift. In our fiducial case, there are hints of the standard deviation of the dust-free distribution changing ($3.4\sigma$), but $<1\sigma$ evidence for the change in the mean of the dust-free distribution. Instead, we find very strong evidence of an evolving quantity of dust with redshift ($>6 \sigma$). Our conservative case finds no evidence for dust-free evolution and a weaker trend in dust evolution $(3\sigma)$, though the weaker dust evolution trend is due to larger statistical uncertainties.} The question of $E_{\rm Dust}$ evolution with the host galaxy mass becomes trickier, however. We have made reference to excluding the first bin in our mass results; this is motivated by the potential of host galaxy mis-association at lower redshift. Our lowest mass bin $M < 8 M_{\rm Star}$ is primarily dominated by low redshift galaxies; those galaxies at high redshift come with significant errors (on or above the order of $1 M_{\rm Star}$). If a SNIa was incorrectly associated to a lower-mass host, it would have the effect of raising the observed $E_{\rm Dust}$ by contamination from the true higher-mass hosts. 

{Rather than a linear model across the host galaxy stellar mass for $E_{\rm Dust}$ and $\beta$, it is possible that the $10^{10}$ mass bin represents a break-point for these parameters. For stellar masses above $10^{10}$, $E_{\rm Dust}$ is consistent with no evolution within $1.7\sigma$; below this mass, the $E_{\rm Dust}$ parameter evolves with mass at a significance of $3.8\sigma$. The opposite occurs with $\beta$: below a mass of $10^{10}$ $\beta$ does not change ($<1\sigma$ change from a flat line), compared to the familiar $3\sigma$ above $10^{10}$. In either case, these results hint at mass-dependent dust properties of SNe Ia.} 

This marks the first time that dust information with this fidelity has been extracted from SNIa lightcurves alone. With the exception of the aforementioned first bin, we find a continuous increase in the columnar density $E_{\rm Dust}$ to a peak and inflection point around $10 M_{\rm Star}$, followed by a decrease. This trend has been found in the literature, notably appearing in \cite{Zahid13} and \cite{Calura17}. \cite{Zahid13} finds an increasing extinction $A_V$  until an inflection point of $10 M_{\rm Star}$, whereby it decreases again. Interestingly, this behaviour holds for all Star Formation Rates. \cite{Calura17} shows increasing dust quantities with host galaxy mass, but only for large elliptical galaxies; spiral galaxies exhibit a flat relationship between mass and dust density. \cite{Rigault18} proposes that this unique shape of the dust property evolution with mass is the byproduct of competing production modes: Star Formation Rate (SFR) driven processes are responsible for increasing the amount of dust up until $\approx 10^{10}$ stellar masses, when star formation stops, and explosion-processes driven by the large galactic stellar mass are the primary driver of dust relations for those galaxies heavier than $\approx 10^{10}$ stellar masses.

\subsection{Evolution of $\beta$ with redshift}

The colour-luminosity coefficient $\beta$ exhibits a strong linear relationship with the host galaxy stellar mass, a relationship that holds regardless of redshift range. This follows the trends seen in \cite{BS20} and \cite{Popovic22}, though those papers only fit in two bins of mass due to computational and statistical constraints, respectively. Previous analyses such as \cite{Sullivan10} and \cite{Gonzalez-Gaitan21} have also found evidence for multiple $\beta$ values, but there has not been evidence that multiple $\beta$ values alone can explain the `mass step'. Unlike \cite{Gonzalez-Gaitan21}, we do not find evidence of a sharp change in $\beta$ at $10 M_{\rm Star}$. Instead, a continuous $\beta$ relationship as a function of the host galaxy $R_v$ is the most probable explanation for this behaviour. 

As a test to ensure that this behaviour of changing $\beta$ with mass is not a sampling issue, we re-fit several downsampled instances of the full data set to check potential biases on $\beta$. We do not find significant biases on $\beta$ ($< 2\sigma$) for samples as small as $\times 20$ smaller than our full data.

In addition, we tested the evolution of $\beta$ with redshift, defining
\begin{equation}\label{eq:betaZ}
\beta(z) = \beta_0 +\beta_1 \times z    
\end{equation}
where $\beta_0$ is the familiar SALT $\beta$ value, and $\beta_1$ is the redshift-dependent term. We find $\beta_0 = 3.25$, but no strong evidence for $\beta$ evolution: $\beta_1 = 0.3 \pm 0.4$. 

\section{Conclusion}\label{sec:Conclusion}

In this paper, we have assembled a sample of $\approx 3000$ SNIa free from magnitude-based selection effects, compiled from ZTF, SDSS, PS1, and DES. This alone marks the largest compilation of SNIa light-curves to-date, demonstrating the  power of the ZTF sample. 

With our `volume-limited' sample and its more conservative counterpart, we have shown strong evidence for evolving host-galaxy properties. The amount of columnar dust density, $E_{\rm Dust}$, changes significantly with both redshift $z$ and the host galaxy mass. This contrasts with the lack of strong evidence for a change in the dust-free SNIa colour distribution, here well-characterised by a Gaussian distribution. 

These results stand independent of potential non-Ia contamination, which we find mitigated by the placement of a Ia-probability cut on the data, and our results persist for the conservative sample, providing strong evidence that these results are not a byproduct of magnitude-limited selection effects. 

Additionally, we find evidence of a continuous and linear relationship between the colour-luminosity coefficient $\beta$ and the host galaxy stellar mass $M_{\rm Star}$. These results together point towards further proof that future standardisation modes for SNIa light-curves must account for the properties of the host galaxy, and that SNe Ia are becoming useful tools for extracting dust properties of their host galaxies.

\section{Acknowledgements}
Based on observations obtained with the Samuel Oschin Telescope 48-inch and the 60-inch Telescope at the Palomar Observatory as part of the Zwicky Transient Facility project. ZTF is supported by the National Science Foundation under Grants No. AST-1440341 and AST-2034437 and a collaboration including current partners Caltech, IPAC, the Weizmann Institute of Science, the Oskar Klein Center at Stockholm University, the University of Maryland, Deutsches Elektronen-Synchrotron and Humboldt University, the TANGO Consortium of Taiwan, the University of Wisconsin at Milwaukee, Trinity College Dublin, Lawrence Livermore National Laboratories, IN2P3, University of Warwick, Ruhr University Bochum, Northwestern University and former partners the University of Washington, Los Alamos National Laboratories, and Lawrence Berkeley National Laboratories. Operations are conducted by COO, IPAC, and UW.
SED Machine is based upon work supported by the National Science Foundation under Grant No. 1106171
The ZTF forced-photometry service was funded under the Heising-Simons Foundation grant \#12540303 (PI: Graham).
This project has received funding from the European Research Council (ERC) under the European Union's Horizon 2020 research and innovation programme (grant agreement n°759194 - USNAC).
This work has been supported by the Agence Nationale de la Recherche of the French government through the program ANR-21-CE31-0016-03.
Y.-L.K. has received funding from the Science and Technology Facilities Council [grant number ST/V000713/1].
This work has been supported by the research project grant “Understanding the Dynamic Universe” funded by the Knut and Alice Wallenberg Foundation under Dnr KAW 2018.0067,  {\em Vetenskapsr\aa det}, the Swedish Research Council, project 2020-03444.
L.G., T.E.M.B acknowledges financial support from the Spanish Ministerio de Ciencia e Innovaci\'on (MCIN) and the Agencia Estatal de Investigaci\'on (AEI) 10.13039/501100011033 under the PID2020-115253GA-I00 HOSTFLOWS project, from Centro Superior de Investigaciones Cient\'ificas (CSIC) under the PIE project 20215AT016 and the program Unidad de Excelencia Mar\'ia de Maeztu CEX2020-001058-M, and from the Departament de Recerca i Universitats de la Generalitat de Catalunya through the 2021-SGR-01270 grant.
Y.-L.K. has received funding from the Science and Technology Facilities Council [grant number ST/V000713/1]. 
UB, GD, JHT are supported by the H2020 European Research Council grant no. 758638

\bibliography{research2.bib}

\begin{thebibliography}{55}
\expandafter\ifx\csname natexlab\endcsname\relax\def\natexlab#1{#1}\fi

\bibitem[{{Amanullah} {et~al.}(2010){Amanullah}, {Lidman}, {Rubin}, {Aldering},
  {Astier}, {Barbary}, {Burns}, {Conley}, {Dawson}, {Deustua}, {Doi}, {Fabbro},
  {Faccioli}, {Fakhouri}, {Folatelli}, {Fruchter}, {Furusawa}, {Garavini},
  {Goldhaber}, {Goobar}, {Groom}, {Hook}, {Howell}, {Kashikawa}, {Kim}, {Knop},
  {Kowalski}, {Linder}, {Meyers}, {Morokuma}, {Nobili}, {Nordin}, {Nugent},
  {{\"O}stman}, {Pain}, {Panagia}, {Perlmutter}, {Raux}, {Ruiz-Lapuente},
  {Spadafora}, {Strovink}, {Suzuki}, {Wang}, {Wood-Vasey}, {Yasuda}, \&
  {Supernova Cosmology Project}}]{Union}
{Amanullah}, R., {Lidman}, C., {Rubin}, D., {et~al.} 2010, \apj, 716, 712

\bibitem[{{Bellm} {et~al.}(2019){Bellm}, {Kulkarni}, {Graham}, {Dekany},
  {Smith}, {Riddle}, {Masci}, {Helou}, {Prince}, {Adams}, {Barbarino},
  {Barlow}, {Bauer}, {Beck}, {Belicki}, {Biswas}, {Blagorodnova}, {Bodewits},
  {Bolin}, {Brinnel}, {Brooke}, {Bue}, {Bulla}, {Burruss}, {Cenko}, {Chang},
  {Connolly}, {Coughlin}, {Cromer}, {Cunningham}, {De}, {Delacroix}, {Desai},
  {Duev}, {Eadie}, {Farnham}, {Feeney}, {Feindt}, {Flynn}, {Franckowiak},
  {Frederick}, {Fremling}, {Gal-Yam}, {Gezari}, {Giomi}, {Goldstein},
  {Golkhou}, {Goobar}, {Groom}, {Hacopians}, {Hale}, {Henning}, {Ho}, {Hover},
  {Howell}, {Hung}, {Huppenkothen}, {Imel}, {Ip}, {Ivezi{\'c}}, {Jackson},
  {Jones}, {Juric}, {Kasliwal}, {Kaspi}, {Kaye}, {Kelley}, {Kowalski},
  {Kramer}, {Kupfer}, {Landry}, {Laher}, {Lee}, {Lin}, {Lin}, {Lunnan},
  {Giomi}, {Mahabal}, {Mao}, {Miller}, {Monkewitz}, {Murphy}, {Ngeow},
  {Nordin}, {Nugent}, {Ofek}, {Patterson}, {Penprase}, {Porter}, {Rauch},
  {Rebbapragada}, {Reiley}, {Rigault}, {Rodriguez}, {van Roestel}, {Rusholme},
  {van Santen}, {Schulze}, {Shupe}, {Singer}, {Soumagnac}, {Stein}, {Surace},
  {Sollerman}, {Szkody}, {Taddia}, {Terek}, {Van Sistine}, {van Velzen},
  {Vestrand}, {Walters}, {Ward}, {Ye}, {Yu}, {Yan}, \& {Zolkower}}]{bellm19}
{Bellm}, E.~C., {Kulkarni}, S.~R., {Graham}, M.~J., {et~al.} 2019, \pasp, 131,
  018002

\bibitem[{{Betoule} {et~al.}(2014){Betoule}, {Kessler}, {Guy}, {Mosher},
  {Hardin}, {Biswas}, {Astier}, {El-Hage}, {Konig}, {Kuhlmann}, {Marriner},
  {Pain}, {Regnault}, {Balland}, {Bassett}, {Brown}, {Campbell}, {Carlberg},
  {Cellier-Holzem}, {Cinabro}, {Conley}, {D'Andrea}, {DePoy}, {Doi}, {Ellis},
  {Fabbro}, {Filippenko}, {Foley}, {Frieman}, {Fouchez}, {Galbany}, {Goobar},
  {Gupta}, {Hill}, {Hlozek}, {Hogan}, {Hook}, {Howell}, {Jha}, {Le Guillou},
  {Leloudas}, {Lidman}, {Marshall}, {M{\"o}ller}, {Mour{\~a}o}, {Neveu},
  {Nichol}, {Olmstead}, {Palanque-Delabrouille}, {Perlmutter}, {Prieto},
  {Pritchet}, {Richmond}, {Riess}, {Ruhlmann-Kleider}, {Sako}, {Schahmaneche},
  {Schneider}, {Smith}, {Sollerman}, {Sullivan}, {Walton}, \&
  {Wheeler}}]{Betoule14}
{Betoule}, M., {Kessler}, R., {Guy}, J., {et~al.} 2014, \aap, 568, A22

\bibitem[{{Blagorodnova} {et~al.}(2018){Blagorodnova}, {Neill}, {Walters},
  {Kulkarni}, {Fremling}, {Ben-Ami}, {Dekany}, {Fucik}, {Konidaris}, {Nash},
  {Ngeow}, {Ofek}, {O' Sullivan}, {Quimby}, {Ritter}, \& {Vyhmeister}}]{SEDm}
{Blagorodnova}, N., {Neill}, J.~D., {Walters}, R., {et~al.} 2018, \pasp, 130,
  035003

\bibitem[{{Briday} {et~al.}(2022){Briday}, {Rigault}, {Graziani}, {Copin},
  {Aldering}, {Amenouche}, {Brinnel}, {Kim}, {Kim}, {Lezmy}, {Nicolas},
  {Nordin}, {Perlmutter}, {Rosnet}, \& {Smith}}]{Briday21}
{Briday}, M., {Rigault}, M., {Graziani}, R., {et~al.} 2022, \aap, 657, A22

\bibitem[{{Brout} \& {Scolnic}(2021)}]{BS20}
{Brout}, D. \& {Scolnic}, D. 2021, \apj, 909, 26

\bibitem[{{Brout} {et~al.}(2022){Brout}, {Scolnic}, {Popovic}, {Riess},
  {Zuntz}, {Kessler}, {Carr}, {Davis}, {Hinton}, {Jones}, {Kenworthy},
  {Peterson}, {Said}, {Taylor}, {Ali}, {Armstrong}, {Charvu}, {Dwomoh},
  {Palmese}, {Qu}, {Rose}, {Stubbs}, {Vincenzi}, {Wood}, {Brown}, {Chen},
  {Chambers}, {Coulter}, {Dai}, {Dimitriadis}, {Filippenko}, {Foley}, {Jha},
  {Kelsey}, {Kirshner}, {M{\"o}ller}, {Muir}, {Nadathur}, {Pan}, {Rest},
  {Rojas-Bravo}, {Sako}, {Siebert}, {Smith}, {Stahl}, \& {Wiseman}}]{Brout22}
{Brout}, D., {Scolnic}, D., {Popovic}, B., {et~al.} 2022, arXiv e-prints,
  arXiv:2202.04077

\bibitem[{{Calura} {et~al.}(2017){Calura}, {Pozzi}, {Cresci}, {Santini},
  {Gruppioni}, {Pozzetti}, {Gilli}, {Matteucci}, \& {Maiolino}}]{Calura17}
{Calura}, F., {Pozzi}, F., {Cresci}, G., {et~al.} 2017, \mnras, 465, 54

\bibitem[{{Chambers} {et~al.}(2016){Chambers}, {Magnier}, {Metcalfe},
  {Flewelling}, {Huber}, {Waters}, {Denneau}, {Draper}, {Farrow}, {Finkbeiner},
  {Holmberg}, {Koppenhoefer}, {Price}, {Saglia}, {Schlafly}, {Smartt},
  {Sweeney}, {Wainscoat}, {Burgett}, {Grav}, {Heasley}, {Hodapp}, {Jedicke},
  {Kaiser}, {Kudritzki}, {Luppino}, {Lupton}, {Monet}, {Morgan}, {Onaka},
  {Stubbs}, {Tonry}, {Banados}, {Bell}, {Bender}, {Bernard}, {Botticella},
  {Casertano}, {Chastel}, {Chen}, {Chen}, {Cole}, {Deacon}, {Frenk},
  {Fitzsimmons}, {Gezari}, {Goessl}, {Goggia}, {Goldman}, {Grebel}, {Hambly},
  {Hasinger}, {Heavens}, {Heckman}, {Henderson}, {Henning}, {Holman}, {Hopp},
  {Ip}, {Isani}, {Keyes}, {Koekemoer}, {Kotak}, {Long}, {Lucey}, {Liu},
  {Martin}, {McLean}, {Morganson}, {Murphy}, {Nieto-Santisteban}, {Norberg},
  {Peacock}, {Pier}, {Postman}, {Primak}, {Rae}, {Rest}, {Riess}, {Riffeser},
  {Rix}, {Roser}, {Schilbach}, {Schultz}, {Scolnic}, {Szalay}, {Seitz},
  {Shiao}, {Small}, {Smith}, {Soderblom}, {Taylor}, {Thakar}, {Thiel},
  {Thilker}, {Urata}, {Valenti}, {Walter}, {Watters}, {Werner}, {White},
  {Wood-Vasey}, \& {Wyse}}]{Chambers16}
{Chambers}, K.~C., {Magnier}, E.~A., {Metcalfe}, N., {et~al.} 2016, ArXiv
  e-prints [\eprint[arXiv]{1612.05560}]

\bibitem[{{Chen} {et~al.}(2022){Chen}, {Scolnic}, {Rozo}, {Rykoff}, {Popovic},
  {Kessler}, {Vincenzi}, {Davis}, {Armstrong}, {Brout}, {Galbany}, {Kelsey},
  {Lidman}, {M{\"o}ller}, {Rose}, {Sako}, {Sullivan}, {Taylor}, {Wiseman},
  {Asorey}, {Carr}, {Conselice}, {Kuehn}, {Lewis}, {Macaulay},
  {Rodriguez-Monroy}, {Tucker}, {Abbott}, {Aguena}, {Allam},
  {Andrade-Oliveira}, {Annis}, {Bacon}, {Bertin}, {Bocquet}, {Brooks}, {Burke},
  {Carnero Rosell}, {Carrasco Kind}, {Carretero}, {Cawthon}, {Costanzi}, {da
  Costa}, {Pereira}, {Desai}, {Diehl}, {Doel}, {Everett}, {Ferrero},
  {Flaugher}, {Friedel}, {Frieman}, {Garc{\'\i}a-Bellido}, {Gatti},
  {Gaztanaga}, {Gruen}, {Hinton}, {Hollowood}, {Honscheid}, {James}, {Lahav},
  {Lima}, {March}, {Menanteau}, {Miquel}, {Morgan}, {Palmese},
  {Paz-Chinch{\'o}n}, {Pieres}, {Plazas Malag{\'o}n}, {Prat}, {Romer},
  {Roodman}, {Sanchez}, {Schubnell}, {Serrano}, {Sevilla-Noarbe}, {Smith},
  {Soares-Santos}, {Suchyta}, {Tarle}, {Thomas}, {To}, {Tucker}, \&
  {Varga}}]{Chen22}
{Chen}, R., {Scolnic}, D., {Rozo}, E., {et~al.} 2022, \apj, 938, 62

\bibitem[{{Childress} {et~al.}(2013){Childress}, {Aldering}, {Antilogus},
  {Aragon}, {Bailey}, {Baltay}, {Bongard}, {Buton}, {Canto}, {Cellier-Holzem},
  {Chotard}, {Copin}, {Fakhouri}, {Gangler}, {Guy}, {Hsiao}, {Kerschhaggl},
  {Kim}, {Kowalski}, {Loken}, {Nugent}, {Paech}, {Pain}, {Pecontal}, {Pereira},
  {Perlmutter}, {Rabinowitz}, {Rigault}, {Runge}, {Scalzo}, {Smadja}, {Tao},
  {Thomas}, {Weaver}, \& {Wu}}]{Childress13}
{Childress}, M., {Aldering}, G., {Antilogus}, P., {et~al.} 2013, \apj, 770, 108

\bibitem[{{Childress} {et~al.}(2014){Childress}, {Wolf}, \&
  {Zahid}}]{Childress14}
{Childress}, M.~J., {Wolf}, C., \& {Zahid}, H.~J. 2014, \mnras, 445, 1898

\bibitem[{{DESI Collaboration} {et~al.}(2024){DESI Collaboration}, {Adame},
  {Aguilar}, {Ahlen}, {Alam}, {Alexander}, {Alvarez}, {Alves}, {Anand},
  {Andrade}, {Armengaud}, {Avila}, {Aviles}, {Awan}, {Bahr-Kalus}, {Bailey},
  {Baltay}, {Bault}, {Behera}, {BenZvi}, {Bera}, {Beutler}, {Bianchi}, {Blake},
  {Blum}, {Brieden}, {Brodzeller}, {Brooks}, {Buckley-Geer}, {Burtin},
  {Calderon}, {Canning}, {Carnero Rosell}, {Cereskaite}, {Cervantes-Cota},
  {Chabanier}, {Chaussidon}, {Chaves-Montero}, {Chen}, {Chen}, {Claybaugh},
  {Cole}, {Cuceu}, {Davis}, {Dawson}, {de la Macorra}, {de Mattia}, {Deiosso},
  {Dey}, {Dey}, {Ding}, {Doel}, {Edelstein}, {Eftekharzadeh}, {Eisenstein},
  {Elliott}, {Fagrelius}, {Fanning}, {Ferraro}, {Ereza}, {Findlay}, {Flaugher},
  {Font-Ribera}, {Forero-S{\'a}nchez}, {Forero-Romero}, {Frenk},
  {Garcia-Quintero}, {Gazta{\~n}aga}, {Gil-Mar{\'\i}n}, {Gontcho},
  {Gonzalez-Morales}, {Gonzalez-Perez}, {Gordon}, {Green}, {Gruen}, {Gsponer},
  {Gutierrez}, {Guy}, {Hadzhiyska}, {Hahn}, {Hanif}, {Herrera-Alcantar},
  {Honscheid}, {Howlett}, {Huterer}, {Ir{\v{s}}i{\v{c}}}, {Ishak}, {Juneau},
  {Kara{\c{c}}ayl{\i}}, {Kehoe}, {Kent}, {Kirkby}, {Kremin}, {Krolewski},
  {Lai}, {Lan}, {Landriau}, {Lang}, {Lasker}, {Le Goff}, {Le Guillou},
  {Leauthaud}, {Levi}, {Li}, {Linder}, {Lodha}, {Magneville}, {Manera},
  {Margala}, {Martini}, {Maus}, {McDonald}, {Medina-Varela}, {Meisner},
  {Mena-Fern{\'a}ndez}, {Miquel}, {Moon}, {Moore}, {Moustakas}, {Mudur},
  {Mueller}, {Mu{\~n}oz-Guti{\'e}rrez}, {Myers}, {Nadathur}, {Napolitano},
  {Neveux}, {Newman}, {Nguyen}, {Nie}, {Niz}, {Noriega}, {Padmanabhan},
  {Paillas}, {Palanque-Delabrouille}, {Pan}, {Penmetsa}, {Percival}, {Pieri},
  {Pinon}, {Poppett}, {Porredon}, {Prada}, {P{\'e}rez-Fern{\'a}ndez},
  {P{\'e}rez-R{\`a}fols}, {Rabinowitz}, {Raichoor}, {Ram{\'\i}rez-P{\'e}rez},
  {Ramirez-Solano}, {Ravoux}, {Rashkovetskyi}, {Rezaie}, {Rich}, {Rocher},
  {Rockosi}, {Roe}, {Rosado-Marin}, {Ross}, {Rossi}, {Ruggeri},
  {Ruhlmann-Kleider}, {Samushia}, {Sanchez}, {Saulder}, {Schlafly}, {Schlegel},
  {Schubnell}, {Seo}, {Shafieloo}, {Sharples}, {Silber}, {Slosar}, {Smith},
  {Sprayberry}, {Tan}, {Tarl{\'e}}, {Taylor}, {Trusov}, {Ure{\~n}a-L{\'o}pez},
  {Vaisakh}, {Valcin}, {Valdes}, {Vargas-Maga{\~n}a}, {Verde}, {Walther},
  {Wang}, {Wang}, {Weaver}, {Weaverdyck}, {Wechsler}, {Weinberg}, {White},
  {Yu}, {Yu}, {Yuan}, {Y{\`e}che}, {Zaborowski}, {Zarrouk}, {Zhang}, {Zhao},
  {Zhao}, {Zhou}, {Zhuang}, \& {Zou}}]{DESIY1}
{DESI Collaboration}, {Adame}, A.~G., {Aguilar}, J., {et~al.} 2024, arXiv
  e-prints, arXiv:2404.03002

\bibitem[{{Foley} {et~al.}(2018){Foley}, {Scolnic}, {Rest}, {Jha}, {Pan},
  {Riess}, {Challis}, {Chambers}, {Coulter}, {Dettman}, {Foley}, {Fox},
  {Huber}, {Jones}, {Kilpatrick}, {Kirshner}, {Schultz}, {Siebert},
  {Flewelling}, {Gibson}, {Magnier}, {Miller}, {Primak}, {Smartt}, {Smith},
  {Wainscoat}, {Waters}, \& {Willman}}]{Foley18}
{Foley}, R.~J., {Scolnic}, D., {Rest}, A., {et~al.} 2018, \mnras, 475, 193

\bibitem[{{Frieman} {et~al.}(2008){Frieman}, {Bassett}, {Becker}, {Choi},
  {Cinabro}, {DeJongh}, {Depoy}, {Dilday}, {Doi}, {Garnavich}, {Hogan},
  {Holtzman}, {Im}, {Jha}, {Kessler}, {Konishi}, {Lampeitl}, {Marriner},
  {Marshall}, {McGinnis}, {Miknaitis}, {Nichol}, {Prieto}, {Riess}, {Richmond},
  {Romani}, {Sako}, {Schneider}, {Smith}, {Takanashi}, {Tokita}, {van der
  Heyden}, {Yasuda}, {Zheng}, {Adelman-McCarthy}, {Annis}, {Assef},
  {Barentine}, {Bender}, {Blandford}, {Boroski}, {Bremer}, {Brewington},
  {Collins}, {Crotts}, {Dembicky}, {Eastman}, {Edge}, {Edmondson}, {Elson},
  {Eyler}, {Filippenko}, {Foley}, {Frank}, {Goobar}, {Gueth}, {Gunn},
  {Harvanek}, {Hopp}, {Ihara}, {Ivezi{\'c}}, {Kahn}, {Kaplan}, {Kent},
  {Ketzeback}, {Kleinman}, {Kollatschny}, {Kron}, {Krzesi{\'n}ski}, {Lamenti},
  {Leloudas}, {Lin}, {Long}, {Lucey}, {Lupton}, {Malanushenko}, {Malanushenko},
  {McMillan}, {Mendez}, {Morgan}, {Morokuma}, {Nitta}, {Ostman}, {Pan},
  {Rockosi}, {Romer}, {Ruiz-Lapuente}, {Saurage}, {Schlesinger}, {Snedden},
  {Sollerman}, {Stoughton}, {Stritzinger}, {Subba Rao}, {Tucker}, {Vaisanen},
  {Watson}, {Watters}, {Wheeler}, {Yanny}, \& {York}}]{Frieman08}
{Frieman}, J.~A., {Bassett}, B., {Becker}, A., {et~al.} 2008, \aj, 135, 338

\bibitem[{{Ginolin} {et~al.}(2024){Ginolin}, {Rigault}, {Copin}, {Popovic},
  {Dimitriadis}, {Goobar}, {Johansson}, {Maguire}, {Nordin}, {Smith}, {Aubert},
  {Barjou-Delayre}, {Burgaz}, {Carreres}, {Dhawan}, {Deckers}, {Feinstein},
  {Fouchez}, {Galbany}, {Ganot}, {de Jaeger}, {Kim}, {Kuhn}, {Lacroix},
  {M{\"u}ller-Bravo}, {Nugent}, {Racine}, {Rosnet}, {Rosselli}, {Ruppin},
  {Sollerman}, {Terwel}, {Townsend}, {Dekany}, {Graham}, {Kasliwal}, {Groom},
  {Purdum}, {Rusholme}, \& {van der Walt}}]{Ginolin24b}
{Ginolin}, M., {Rigault}, M., {Copin}, Y., {et~al.} 2024, arXiv e-prints,
  arXiv:2406.02072

\bibitem[{{Gonz{\'a}lez-Gait{\'a}n} {et~al.}(2021){Gonz{\'a}lez-Gait{\'a}n},
  {de Jaeger}, {Galbany}, {Mour{\~a}o}, {Paulino-Afonso}, \&
  {Filippenko}}]{Gonzalez-Gaitan21}
{Gonz{\'a}lez-Gait{\'a}n}, S., {de Jaeger}, T., {Galbany}, L., {et~al.} 2021,
  \mnras, 508, 4656

\bibitem[{{Graham} {et~al.}(2019){Graham}, {Kulkarni}, {Bellm}, {Adams},
  {Barbarino}, {Blagorodnova}, {Bodewits}, {Bolin}, {Brady}, {Cenko}, {Chang},
  {Coughlin}, {De}, {Eadie}, {Farnham}, {Feindt}, {Franckowiak}, {Fremling},
  {Gezari}, {Ghosh}, {Goldstein}, {Golkhou}, {Goobar}, {Ho}, {Huppenkothen},
  {Ivezi{\'c}}, {Jones}, {Juric}, {Kaplan}, {Kasliwal}, {Kelley}, {Kupfer},
  {Lee}, {Lin}, {Lunnan}, {Mahabal}, {Miller}, {Ngeow}, {Nugent}, {Ofek},
  {Prince}, {Rauch}, {van Roestel}, {Schulze}, {Singer}, {Sollerman}, {Taddia},
  {Yan}, {Ye}, {Yu}, {Barlow}, {Bauer}, {Beck}, {Belicki}, {Biswas}, {Brinnel},
  {Brooke}, {Bue}, {Bulla}, {Burruss}, {Connolly}, {Cromer}, {Cunningham},
  {Dekany}, {Delacroix}, {Desai}, {Duev}, {Feeney}, {Flynn}, {Frederick},
  {Gal-Yam}, {Giomi}, {Groom}, {Hacopians}, {Hale}, {Helou}, {Henning},
  {Hover}, {Hillenbrand}, {Howell}, {Hung}, {Imel}, {Ip}, {Jackson}, {Kaspi},
  {Kaye}, {Kowalski}, {Kramer}, {Kuhn}, {Landry}, {Laher}, {Mao}, {Masci},
  {Monkewitz}, {Murphy}, {Nordin}, {Patterson}, {Penprase}, {Porter},
  {Rebbapragada}, {Reiley}, {Riddle}, {Rigault}, {Rodriguez}, {Rusholme}, {van
  Santen}, {Shupe}, {Smith}, {Soumagnac}, {Stein}, {Surace}, {Szkody}, {Terek},
  {Van Sistine}, {van Velzen}, {Vestrand}, {Walters}, {Ward}, {Zhang}, \&
  {Zolkower}}]{graham19}
{Graham}, M.~J., {Kulkarni}, S.~R., {Bellm}, E.~C., {et~al.} 2019, \pasp, 131,
  078001

\bibitem[{{Guy} {et~al.}(2010){Guy}, {Sullivan}, {Conley}, {Regnault},
  {Astier}, {Balland}, {Basa}, {Carlberg}, {Fouchez}, {Hardin}, {Hook},
  {Howell}, {Pain}, {Palanque-Delabrouille}, {Perrett}, {Pritchet}, {Rich},
  {Ruhlmann-Kleider}, {Balam}, {Baumont}, {Ellis}, {Fabbro}, {Fakhouri},
  {Fourmanoit}, {Gonz{\'a}lez-Gait{\'a}n}, {Graham}, {Hsiao}, {Kronborg},
  {Lidman}, {Mourao}, {Perlmutter}, {Ripoche}, {Suzuki}, \& {Walker}}]{Guy10}
{Guy}, J., {Sullivan}, M., {Conley}, A., {et~al.} 2010, \aap, 523, A7

\bibitem[{{Hamuy} {et~al.}(1996){Hamuy}, {Phillips}, {Suntzeff}, {Schommer},
  {Maza}, \& {Aviles}}]{Hamuy96}
{Hamuy}, M., {Phillips}, M.~M., {Suntzeff}, N.~B., {et~al.} 1996, \aj, 112,
  2391

\bibitem[{{Hicken} {et~al.}(2009){Hicken}, {Challis}, {Jha}, {Kirshner},
  {Matheson}, {Modjaz}, {Rest}, {Wood-Vasey}, {Bakos}, {Barton}, {Berlind},
  {Bragg}, {Brice{\~n}o}, {Brown}, {Caldwell}, {Calkins}, {Cho}, {Ciupik},
  {Contreras}, {Dendy}, {Dosaj}, {Durham}, {Eriksen}, {Esquerdo}, {Everett},
  {Falco}, {Fernandez}, {Gaba}, {Garnavich}, {Graves}, {Green}, {Groner},
  {Hergenrother}, {Holman}, {Hradecky}, {Huchra}, {Hutchison}, {Jerius},
  {Jordan}, {Kilgard}, {Krauss}, {Luhman}, {Macri}, {Marrone}, {McDowell},
  {McIntosh}, {McNamara}, {Megeath}, {Mochejska}, {Munoz}, {Muzerolle},
  {Naranjo}, {Narayan}, {Pahre}, {Peters}, {Peterson}, {Rines}, {Ripman},
  {Roussanova}, {Schild}, {Sicilia-Aguilar}, {Sokoloski}, {Smalley}, {Smith},
  {Spahr}, {Stanek}, {Barmby}, {Blondin}, {Stubbs}, {Szentgyorgyi}, {Torres},
  {Vaz}, {Vikhlinin}, {Wang}, {Westover}, {Woods}, \& {Zhao}}]{Hicken09a}
{Hicken}, M., {Challis}, P., {Jha}, S., {et~al.} 2009, \apj, 700, 331

\bibitem[{{Howell} {et~al.}(2007){Howell}, {Sullivan}, {Conley}, \&
  {Carlberg}}]{Howell07}
{Howell}, D.~A., {Sullivan}, M., {Conley}, A., \& {Carlberg}, R. 2007, \apjl,
  667, L37

\bibitem[{{James} \& {Roos}(1975)}]{MINUIT}
{James}, F. \& {Roos}, M. 1975, Computer Physics Communications, 10, 343

\bibitem[{Johansson {et~al.}(2021)Johansson, Cenko, Fox, Dhawan, Goobar,
  Stanishev, Butler, Lee, Watson, Fremling, Kasliwal, Nugent, Petrushevska,
  Sollerman, Yan, Burke, Hosseinzadeh, Howell, McCully, \&
  Valenti}]{Johansson20}
Johansson, J., Cenko, S.~B., Fox, O.~D., {et~al.} 2021, The Astrophysical
  Journal, 923, 237

\bibitem[{{Jones} {et~al.}(2018){Jones}, {Riess}, {Scolnic}, {Pan}, {Johnson},
  {Coulter}, {Dettman}, {Foley}, {Foley}, {Huber}, {Jha}, {Kilpatrick},
  {Kirshner}, {Rest}, {Schultz}, \& {Siebert}}]{Jones18}
{Jones}, D.~O., {Riess}, A.~G., {Scolnic}, D.~M., {et~al.} 2018, \apj, 867, 108

\bibitem[{{Kelly} {et~al.}(2010){Kelly}, {Hicken}, {Burke}, {Mandel}, \&
  {Kirshner}}]{Kelly10}
{Kelly}, P.~L., {Hicken}, M., {Burke}, D.~L., {Mandel}, K.~S., \& {Kirshner},
  R.~P. 2010, \apj, 715, 743

\bibitem[{{Kelsey} {et~al.}(2022){Kelsey}, {Sullivan}, {Wiseman}, {Armstrong},
  {Chen}, {Brout}, {Davis}, {Dixon}, {Frohmaier}, {Galbany}, {Graur},
  {Kessler}, {Lidman}, {M{\"o}ller}, {Popovic}, {Rose}, {Scolnic}, {Smith},
  {Vincenzi}, {Abbott}, {Aguena}, {Allam}, {Alves}, {Annis}, {Bacon}, {Bertin},
  {Bocquet}, {Brooks}, {Burke}, {Carnero Rosell}, {Carrasco Kind}, {Carretero},
  {Costanzi}, {da Costa}, {Pereira}, {Desai}, {Diehl}, {Everett}, {Ferrero},
  {Frieman}, {Garc{\'\i}a-Bellido}, {Gruen}, {Gruendl}, {Gschwend},
  {Gutierrez}, {Hinton}, {Hollowood}, {Honscheid}, {James}, {Kuehn},
  {Kuropatkin}, {Lewis}, {Mena-Fern{\'a}ndez}, {Miquel}, {Palmese},
  {Paz-Chinch{\'o}n}, {Pieres}, {Plazas Malag{\'o}n}, {Raveri},
  {Rodriguez-Monroy}, {Romer}, {Sanchez}, {Scarpine}, {Schubnell},
  {Sevilla-Noarbe}, {Suchyta}, {Swanson}, {Tarle}, {Tucker}, \&
  {Weaverdyck}}]{Kelsey22}
{Kelsey}, L., {Sullivan}, M., {Wiseman}, P., {et~al.} 2022, \mnras
  [\eprint[arXiv]{2208.01357}]

\bibitem[{{Kessler} \& {Scolnic}(2017)}]{Kessler16}
{Kessler}, R. \& {Scolnic}, D. 2017, \apj, 836, 56

\bibitem[{{Krisciunas} {et~al.}(2017){Krisciunas}, {Contreras}, {Burns},
  {Phillips}, {Stritzinger}, {Morrell}, {Hamuy}, {Anais}, {Boldt}, {Busta},
  {Campillay}, {Castell{\'o}n}, {Folatelli}, {Freedman}, {Gonz{\'a}lez},
  {Hsiao}, {Krzeminski}, {Persson}, {Roth}, {Salgado}, {Ser{\'o}n}, {Suntzeff},
  {Torres}, {Filippenko}, {Li}, {Madore}, {DePoy}, {Marshall}, {Rheault}, \&
  {Villanueva}}]{Krisciunas17}
{Krisciunas}, K., {Contreras}, C., {Burns}, C.~R., {et~al.} 2017, \aj, 154, 211

\bibitem[{{Kroupa}(2001)}]{Kroupa01}
{Kroupa}, P. 2001, \mnras, 322, 231

\bibitem[{{Lampeitl} {et~al.}(2010){Lampeitl}, {Smith}, {Nichol}, {Bassett},
  {Cinabro}, {Dilday}, {Foley}, {Frieman}, {Garnavich}, {Goobar}, {Im}, {Jha},
  {Marriner}, {Miquel}, {Nordin}, {{\"O}stman}, {Riess}, {Sako}, {Schneider},
  {Sollerman}, \& {Stritzinger}}]{Lampeitl10}
{Lampeitl}, H., {Smith}, M., {Nichol}, R.~C., {et~al.} 2010, \apj, 722, 566

\bibitem[{{Mandel} {et~al.}(2017){Mandel}, {Scolnic}, {Shariff}, {Foley}, \&
  {Kirshner}}]{Mandel17}
{Mandel}, K.~S., {Scolnic}, D.~M., {Shariff}, H., {Foley}, R.~J., \&
  {Kirshner}, R.~P. 2017, \apj, 842, 93

\bibitem[{{Masci} {et~al.}(2019){Masci}, {Laher}, {Rusholme}, {Shupe}, {Groom},
  {Surace}, {Jackson}, {Monkewitz}, {Beck}, {Flynn}, {Terek}, {Landry},
  {Hacopians}, {Desai}, {Howell}, {Brooke}, {Imel}, {Wachter}, {Ye}, {Lin},
  {Cenko}, {Cunningham}, {Rebbapragada}, {Bue}, {Miller}, {Mahabal}, {Bellm},
  {Patterson}, {Juri{\'c}}, {Golkhou}, {Ofek}, {Walters}, {Graham}, {Kasliwal},
  {Dekany}, {Kupfer}, {Burdge}, {Cannella}, {Barlow}, {Van Sistine}, {Giomi},
  {Fremling}, {Blagorodnova}, {Levitan}, {Riddle}, {Smith}, {Helou}, {Prince},
  \& {Kulkarni}}]{masci19}
{Masci}, F.~J., {Laher}, R.~R., {Rusholme}, B., {et~al.} 2019, \pasp, 131,
  018003

\bibitem[{{M{\"o}ller} \& {de Boissi{\`e}re}(2019)}]{Moller19}
{M{\"o}ller}, A. \& {de Boissi{\`e}re}, T. 2019, arXiv e-prints,
  arXiv:1901.06384

\bibitem[{{Nicolas} {et~al.}(2020){Nicolas}, {Rigault}, {Copin}, {Graziani},
  {Aldering}, {Briday}, {Nordin}, {Kim}, \& {Perlmutter}}]{Nicolas21}
{Nicolas}, N., {Rigault}, M., {Copin}, Y., {et~al.} 2020, arXiv e-prints,
  arXiv:2005.09441

\bibitem[{{Paulino-Afonso} {et~al.}(2022){Paulino-Afonso},
  {Gonz{\'a}lez-Gait{\'a}n}, {Galbany}, {Maria Mour{\~a}o}, {Angus}, {Smith},
  {Anderson}, {Lyman}, {Kuncarayakti}, \& {Rodrigues}}]{Alfonso22}
{Paulino-Afonso}, A., {Gonz{\'a}lez-Gait{\'a}n}, S., {Galbany}, L., {et~al.}
  2022, \aap, 662, A86

\bibitem[{{Perlmutter} {et~al.}(1999){Perlmutter}, {Aldering}, {Goldhaber},
  {Knop}, {Nugent}, {Castro}, {Deustua}, {Fabbro}, {Goobar}, {Groom}, {Hook},
  {Kim}, {Kim}, {Lee}, {Nunes}, {Pain}, {Pennypacker}, {Quimby}, {Lidman},
  {Ellis}, {Irwin}, {McMahon}, {Ruiz-Lapuente}, {Walton}, {Schaefer}, {Boyle},
  {Filippenko}, {Matheson}, {Fruchter}, {Panagia}, {Newberg}, {Couch}, \&
  {Project}}]{Perlmutter99}
{Perlmutter}, S., {Aldering}, G., {Goldhaber}, G., {et~al.} 1999, \apj, 517,
  565

\bibitem[{{Popovic} {et~al.}(2021{\natexlab{a}}){Popovic}, {Brout}, {Kessler},
  \& {Scolnic}}]{Popovic22}
{Popovic}, B., {Brout}, D., {Kessler}, R., \& {Scolnic}, D. 2021{\natexlab{a}},
  arXiv e-prints, arXiv:2112.04456

\bibitem[{{Popovic} {et~al.}(2021{\natexlab{b}}){Popovic}, {Brout}, {Kessler},
  {Scolnic}, \& {Lu}}]{Popovic21a}
{Popovic}, B., {Brout}, D., {Kessler}, R., {Scolnic}, D., \& {Lu}, L.
  2021{\natexlab{b}}, \apj, 913, 49

\bibitem[{{Popovic} {et~al.}(2019){Popovic}, {Scolnic}, \&
  {Kessler}}]{Popovic19}
{Popovic}, B., {Scolnic}, D., \& {Kessler}, R. 2019, arXiv e-prints,
  arXiv:1910.05228

\bibitem[{{Popovic} {et~al.}(2023){Popovic}, {Scolnic}, {Vincenzi}, {Sullivan},
  {Brout}, {Sanchez}, {Chen}, {Patel}, {Peterson}, {Kessler}, {Kelsey},
  {Bailey}, {Wiseman}, \& {Toy}}]{Amalgame}
{Popovic}, B., {Scolnic}, D., {Vincenzi}, M., {et~al.} 2023, arXiv e-prints,
  arXiv:2309.05654

\bibitem[{{Riess} {et~al.}(1998){Riess}, {Filippenko}, {Challis},
  {Clocchiatti}, {Diercks}, {Garnavich}, {Gilliland}, {Hogan}, {Jha},
  {Kirshner}, {Leibundgut}, {Phillips}, {Reiss}, {Schmidt}, {Schommer},
  {Smith}, {Spyromilio}, {Stubbs}, {Suntzeff}, \& {Tonry}}]{Riess98}
{Riess}, A.~G., {Filippenko}, A.~V., {Challis}, P., {et~al.} 1998, \aj, 116,
  1009

\bibitem[{{Rigault} {et~al.}(2020){Rigault}, {Brinnel}, {Aldering},
  {Antilogus}, {Aragon}, {Bailey}, {Baltay}, {Barbary}, {Bongard}, {Boone},
  {Buton}, {Childress}, {Chotard}, {Copin}, {Dixon}, {Fagrelius}, {Feindt},
  {Fouchez}, {Gangler}, {Hayden}, {Hillebrandt}, {Howell}, {Kim}, {Kowalski},
  {Kuesters}, {Leget}, {Lombardo}, {Lin}, {Nordin}, {Pain}, {Pecontal},
  {Pereira}, {Perlmutter}, {Rabinowitz}, {Runge}, {Rubin}, {Saunders},
  {Smadja}, {Sofiatti}, {Suzuki}, {Taubenberger}, {Tao}, \&
  {Thomas}}]{Rigault18}
{Rigault}, M., {Brinnel}, V., {Aldering}, G., {et~al.} 2020, \aap, 644, A176

\bibitem[{{Rigault} {et~al.}(2013){Rigault}, {Copin}, {Aldering}, {Antilogus},
  {Aragon}, {Bailey}, {Baltay}, {Bongard}, {Buton}, {Canto}, {Cellier-Holzem},
  {Childress}, {Chotard}, {Fakhouri}, {Feindt}, {Fleury}, {Gangler},
  {Greskovic}, {Guy}, {Kim}, {Kowalski}, {Lombardo}, {Nordin}, {Nugent},
  {Pain}, {P{\'e}contal}, {Pereira}, {Perlmutter}, {Rabinowitz}, {Runge},
  {Saunders}, {Scalzo}, {Smadja}, {Tao}, {Thomas}, \& {Weaver}}]{Rigault13}
{Rigault}, M., {Copin}, Y., {Aldering}, G., {et~al.} 2013, \aap, 560, A66

\bibitem[{{Rigault} {et~al.}(2019){Rigault}, {Neill}, {Blagorodnova}, {Dugas},
  {Feeney}, {Walters}, {Brinnel}, {Copin}, {Fremling}, {Nordin}, \&
  {Sollerman}}]{IFU}
{Rigault}, M., {Neill}, J.~D., {Blagorodnova}, N., {et~al.} 2019, \aap, 627,
  A115

\bibitem[{{Sako} {et~al.}(2018){Sako}, {Bassett}, {Becker}, {Brown},
  {Campbell}, {Wolf}, {Cinabro}, {D'Andrea}, {Dawson}, {DeJongh}, {Depoy},
  {Dilday}, {Doi}, {Filippenko}, {Fischer}, {Foley}, {Frieman}, {Galbany},
  {Garnavich}, {Goobar}, {Gupta}, {Hill}, {Hayden}, {Hlozek}, {Holtzman},
  {Hopp}, {Jha}, {Kessler}, {Kollatschny}, {Leloudas}, {Marriner}, {Marshall},
  {Miquel}, {Morokuma}, {Mosher}, {Nichol}, {Nordin}, {Olmstead}, {{\"O}stman},
  {Prieto}, {Richmond}, {Romani}, {Sollerman}, {Stritzinger}, {Schneider},
  {Smith}, {Wheeler}, {Yasuda}, \& {Zheng}}]{Sako18}
{Sako}, M., {Bassett}, B., {Becker}, A.~C., {et~al.} 2018, Publications of the
  Astronomical Society of the Pacific, 130, 064002

\bibitem[{{Scolnic} {et~al.}(2018){Scolnic}, {Kessler}, {Brout},
  {Cowperthwaite}, {Soares-Santos}, {Annis}, {Herner}, {Chen}, {Sako},
  {Doctor}, {Butler}, {Palmese}, {Diehl}, {Frieman}, {Holz}, {Berger},
  {Chornock}, {Villar}, {Nicholl}, {Biswas}, {Hounsell}, {Foley}, {Metzger},
  {Rest}, {Garc{\'{\i}}a-Bellido}, {M{\"o}ller}, {Nugent}, {Abbott}, {Abdalla},
  {Allam}, {Bechtol}, {Benoit-L{\'e}vy}, {Bertin}, {Brooks}, {Buckley-Geer},
  {Carnero Rosell}, {Carrasco Kind}, {Carretero}, {Castander}, {Cunha},
  {D{'}Andrea}, {da Costa}, {Davis}, {Doel}, {Drlica-Wagner}, {Eifler},
  {Flaugher}, {Fosalba}, {Gaztanaga}, {Gerdes}, {Gruen}, {Gruendl}, {Gschwend},
  {Gutierrez}, {Hartley}, {Honscheid}, {James}, {Johnson}, {Johnson}, {Krause},
  {Kuehn}, {Kuhlmann}, {Lahav}, {Li}, {Lima}, {Maia}, {March}, {Marshall},
  {Menanteau}, {Miquel}, {Neilsen}, {Plazas}, {Sanchez}, {Scarpine},
  {Schubnell}, {Sevilla-Noarbe}, {Smith}, {Smith}, {Sobreira}, {Suchyta},
  {Swanson}, {Tarle}, {Thomas}, {Tucker}, {Walker}, \& {DES
  Collaboration}}]{Scolnic18}
{Scolnic}, D., {Kessler}, R., {Brout}, D., {et~al.} 2018, \apjl, 852, L3

\bibitem[{{Sullivan} {et~al.}(2010){Sullivan}, {Conley}, {Howell}, {Neill},
  {Astier}, {Balland}, {Basa}, {Carlberg}, {Fouchez}, {Guy}, {Hardin}, {Hook},
  {Pain}, {Palanque-Delabrouille}, {Perrett}, {Pritchet}, {Regnault}, {Rich},
  {Ruhlmann-Kleider}, {Baumont}, {Hsiao}, {Kronborg}, {Lidman}, {Perlmutter},
  \& {Walker}}]{Sullivan10}
{Sullivan}, M., {Conley}, A., {Howell}, D.~A., {et~al.} 2010, \mnras, 406, 782

\bibitem[{{Sullivan} {et~al.}(2006){Sullivan}, {Le Borgne}, {Pritchet},
  {Hodsman}, {Neill}, {Howell}, {Carlberg}, {Astier}, {Aubourg}, {Balam},
  {Basa}, {Conley}, {Fabbro}, {Fouchez}, {Guy}, {Hook}, {Pain},
  {Palanque-Delabrouille}, {Perrett}, {Regnault}, {Rich}, {Taillet}, {Baumont},
  {Bronder}, {Ellis}, {Filiol}, {Lusset}, {Perlmutter}, {Ripoche}, \&
  {Tao}}]{Sullivan06}
{Sullivan}, M., {Le Borgne}, D., {Pritchet}, C.~J., {et~al.} 2006, \apj, 648,
  868

\bibitem[{{Taylor} {et~al.}(2021){Taylor}, {Lidman}, {Tucker}, {Brout},
  {Hinton}, \& {Kessler}}]{Taylor21}
{Taylor}, G., {Lidman}, C., {Tucker}, B.~E., {et~al.} 2021, \mnras, 504, 4111

\bibitem[{{Tripp}(1998)}]{Tripp98}
{Tripp}, R. 1998, \aap, 331, 815

\bibitem[{{Uddin} {et~al.}(2017){Uddin}, {Mould}, {Lidman}, {Ruhlmann-Kleider},
  \& {Zhang}}]{Uddin17}
{Uddin}, S.~A., {Mould}, J., {Lidman}, C., {Ruhlmann-Kleider}, V., \& {Zhang},
  B.~R. 2017, \apj, 848, 56

\bibitem[{{Vincenzi} {et~al.}(in prep.){Vincenzi}, {Brout}, {Armstrong},
  {Popovic}, {Taylor}, {Acevedo}, R., {Chen}, T., J., C., S., L., R., A., {Qu},
  {Sako}, {Sanchez}, {Scolnic}, {Smith}, {Sullivan}, {Wiseman}, {DES WG}, \&
  {DES Builders}}]{DES5YR}
{Vincenzi}, M., {Brout}, D., {Armstrong}, P., {et~al.} in prep.

\bibitem[{{Wiseman} {et~al.}(2022){Wiseman}, {Vincenzi}, {Sullivan}, {Kelsey},
  {Popovic}, {Rose}, {Brout}, {Davis}, {Frohmaier}, {Galbany}, {Lidman},
  {M{\"o}ller}, {Scolnic}, {Smith}, {Aguena}, {Allam}, {Andrade-Oliveira},
  {Annis}, {Bertin}, {Bocquet}, {Brooks}, {Burke}, {Carnero Rosell}, {Carrasco
  Kind}, {Carretero}, {Castander}, {Costanzi}, {Pereira}, {Desai}, {Diehl},
  {Doel}, {Everett}, {Ferrero}, {Friedel}, {Frieman}, {Garc{\'\i}a-Bellido},
  {Gatti}, {Gaztanaga}, {Gruen}, {Gschwend}, {Gutierrez}, {Hinton},
  {Hollowood}, {Honscheid}, {James}, {March}, {Menanteau}, {Miquel}, {Morgan},
  {Palmese}, {Paz-Chinch{\'o}n}, {Pieres}, {Plazas Malag{\'o}n}, {Romer},
  {Sanchez}, {Scarpine}, {Sevilla-Noarbe}, {Soares-Santos}, {Suchyta}, {Tarle},
  {To}, {Varga}, \& {DES Collaboration}}]{Wiseman22}
{Wiseman}, P., {Vincenzi}, M., {Sullivan}, M., {et~al.} 2022, \mnras, 515, 4587

\bibitem[{{Zahid} {et~al.}(2013){Zahid}, {Yates}, {Kewley}, \&
  {Kudritzki}}]{Zahid13}
{Zahid}, H.~J., {Yates}, R.~M., {Kewley}, L.~J., \& {Kudritzki}, R.~P. 2013,
  \apj, 763, 92

\end{thebibliography}

\bibliographystyle{aa.bst}

\end{document}